\pdfoutput=1

\documentclass{article}

\usepackage{amssymb}
\usepackage{graphicx}
\usepackage{lineno}
\usepackage{lscape}
\usepackage{subfig}

\usepackage[colorlinks=true,linkcolor=black, citecolor=blue, urlcolor=blue]{hyperref}
\usepackage{changes}

\usepackage{amsmath}
\usepackage{algorithm}
\usepackage[noend]{algpseudocode}
\usepackage{relsize}

\newcommand{\pdp}{%
	\mathrel{%
		\vcenter{\offinterlineskip
			\ialign{##\cr$^{P_i}$\cr\noalign{\kern-1.5pt}$\rightharpoondown$\cr}%
		}%
	}%
}

\newcommand{\pdc}{%
	\mathrel{%
		\vcenter{\offinterlineskip
			\ialign{##\cr$^{C_i}$\cr\noalign{\kern-1.5pt}$\rightharpoondown$\cr}%
		}%
	}%
}

\newcounter{matriz}
\newenvironment{matriz}{\refstepcounter{matriz}\equation}{\tag{Rule \thematriz}\endequation}

\usepackage{setspace}

\begin{document}
	
	\title{Consistency models in distributed systems:\\ A survey on definitions, disciplines, challenges and applications}
	
	\author{Hesam Nejati Sharif Aldin \footnote{Corresponding author.}\\
		Department of Computer Engineering,\\
		Mashhad Branch, Islamic Azad University,\\
		Mashhad, Iran\\
		\texttt{hesam.nejati@mshdiau.ac.ir}\\
		\and Hossein Deldari \\
		Department of Computer Engineering, \\
		Ferdowsi University of Mashhad, Iran\\
		\texttt{hd@ferdowsi.um.ac.ir}\\
		\and
		Mohammad Hossein Moattar \\
		Department of Computer Engineering,\\
		Mashhad Branch, Islamic Azad University,\\
		Mashhad, Iran\\
		\texttt{moattar@mshdiau.ac.ir} \\
		\and
		Mostafa Razavi Ghods \\
		Department of Computer Engineering,\\
		Mashhad Branch, Islamic Azad University,\\
		Mashhad, Iran\\
		\texttt{mrazavi@mshdiau.ac.ir} \\
	}
	
	
	\maketitle

%
%
%
%
%
%
%
%
%
		\begin{abstract}
			The replication mechanism resolves some challenges with big data such as data durability, data access, and fault tolerance. Yet, replication itself gives birth to another challenge known as the consistency in distributed systems. Scalability and availability are the challenging criteria on which the replication is based upon in distributed systems which themselves require the consistency. Consistency in distributed computing systems has been employed in three different applicable fields, such as system architecture, distributed database, and distributed systems. Consistency models based on their applicability could be sorted from strong to weak. Our goal is to propose a novel viewpoint to different consistency models utilized in the distributed systems. This research proposes two different categories of consistency models. Initially, consistency models are categorized into three groups of data-centric, client-centric and hybrid models. Each of which is then grouped into three subcategories of traditional, extended, and novel consistency models. Consequently, the concepts and procedures are expressed in mathematical terms, which are introduced in order to present our models’ behavior without implementation. Moreover, we have surveyed different aspects of challenges with respect to the consistency i.e., availability, scalability, security, fault tolerance, latency, violation, and staleness, out of which the two latter i.e. violation and staleness, play the most pivotal roles in terms of consistency and trade-off balancing. Finally, the contribution extent of each of the consistency models and the growing need for them in distributed systems are investigated.
		\end{abstract}
%
%
%
	
	\section{Introduction}\label{sec.Intro}
	Nowadays we are faced with an enormous amount of data which give birth to concepts like big data.
	Big data, is a sort of too big, massive and extensive data \cite{bouge2013managing}. These floods of digital data are generated from variant sources such as sensors, digitizers, scanners, cell phones, the Internet, emails, and social networks. The diversity of these data covers the pictures, films, sounds and a combination of each of them \cite{Yang2017}. The evolution of technology and human knowledge about data is formed by analyzing its development from the static traditional to the rapid form with respect to some of the characteristics of big data. The common concept between the researchers in that the big data infers to the set of data with characteristic of volume, variety, and velocity \cite{laney20013d}. Amongst, some the researchers and specialists refer to some other properties like the value \cite{Yang2017,Hashem2015,KacfahEmani2015}, veracity \cite{Yang2017,KacfahEmani2015}, variability \cite{bouge2013managing,KacfahEmani2015}, and complexity \cite{bouge2013managing}.
	
	The challenges of these data is the problem that the data and database technicians are come fronted for many years. What is learned through these years, is the way to face the challenges of the big data in small scales. Challenges like the data durability, data access, data availability, and fault tolerance \cite{bouge2013managing} which are generally solvable with the replication mechanism. Replication is a crucial challenge in the big data. The guarantee of the consistency is the challenge that the replication mechanism brings about.
	 
	 The replication and caching process are used as techniques to make the scalability achievable. Data replication \cite{Steen2009} is meant to generate a number of indistinguishable copies of the original i.e., the replicas. One of the major problems is keeping the replica consistent, though the interaction between the replica is inevitable. In other words, the replication in the distributed systems demands the consistency's guarantee and is one of the major problems in large scale storage systems \cite{Lee2015}. From the viewpoint of the researchers, consistency is meant to have multiple processes have access to common data. At this point, consistency means that each process have knowledge of the other processes have access to the resource (whether they read or write) and also know what to expect. The main reason of the replication is the concurrent access to the replica \cite{Steen2009}. Consistency is a part of the system behavior, in order to make  concurrent execution or system failure predictable \cite{Lee2015}. The tradeoff between the performance and consistency has made the researchers to look for consistency policies like consistency level and technology. The consistency policy is based on the principle of what should be written or read. The placement policy shows the demand-caching, prefetching, push-caching and full replication that the nodes store which data from which local copy. The technology policy, like the client-server is hierarchical or ad hoc and the directories along which the communicative data is streamed are defined \cite{belaramani2006practi}. Up to now, we have introduced a variety of consistency models which have been used in the transaction less distributed systems \cite{Viotti2016}. However, the proceeds of the consistency models alter based on the application in which they are employed. Different criteria and services have been considered for data-sharing in the distributed systems, out of which, the five most important criteria are \cite{susarla2003composable}:
	
	\begin{itemize}
		\item Concurrency: the degree at which the conflicting read/write access is tolerated.
		
		\item Consistency: preservation of the update dependency and stale read data tolerance.
		
		\item Availability: the access method to the replica is their absence.
		
		\item Visibility: How should we have a global view, when the local changes have been applied on the replicated data.
		
		\item Isolation: when the remote updating must be observed locally.
	\end{itemize} 

\begin{figure}
	\includegraphics[width=\columnwidth]{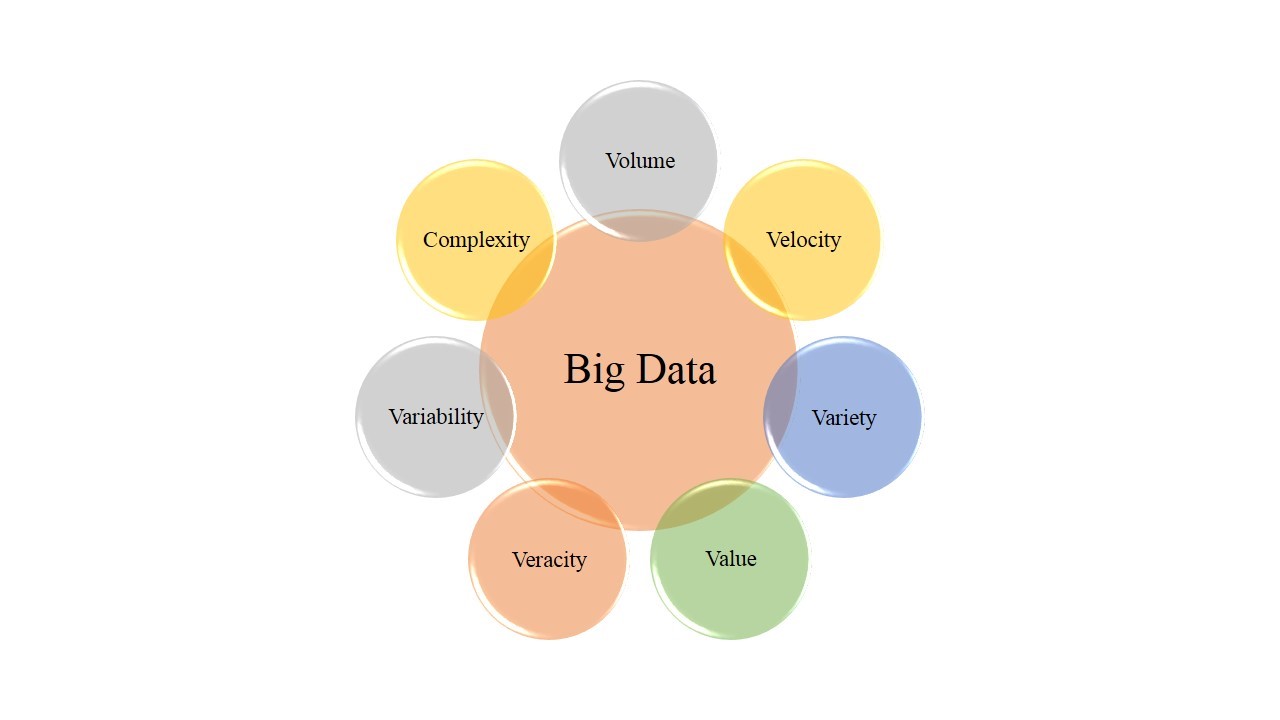}\\
	\centering{
		\caption{The characteristics of the big data.}
		\label{fig.BigData}
	}
\end{figure}
	
\begin{figure}
	\includegraphics[width=\textwidth, scale = 1]{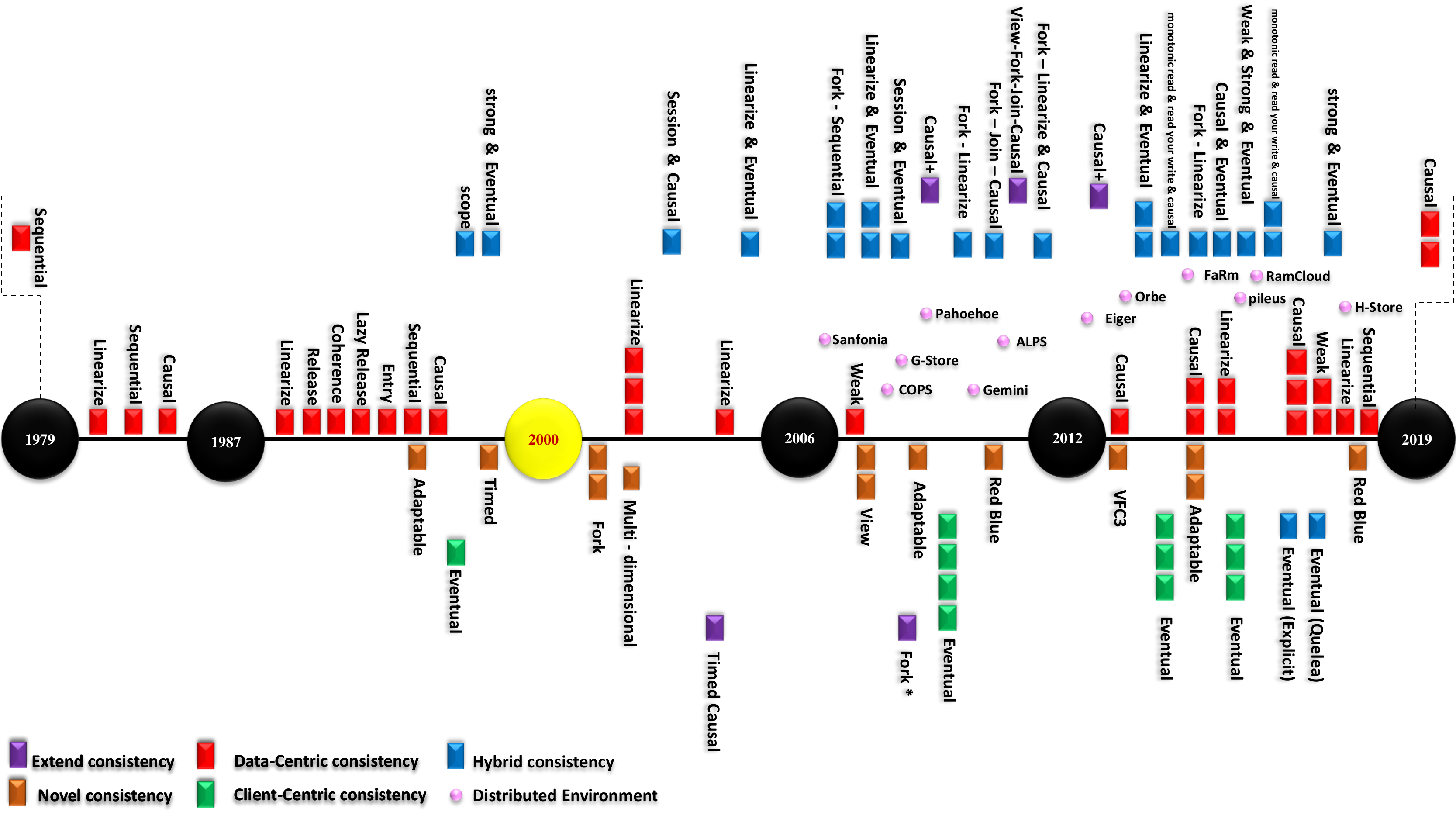}\\
	\centering{
		\caption{Classification of different consistency models used in distributed systems.}
		\label{fig.timeLine}
	}
\end{figure}

	The mentioned criteria, show the different aspects of the consistency application requirements. With different combination of these criteria, a majority of consistency semantics for reading or updating the shared data will be emerged \cite{susarla2003composable}. Totally, the data consistency can be grouped into to two categories of the data-centric and client-centric. Data-centric consistency is the model where there exists a contract between the data-center and the processes. This models says that if the processes agree to obey certain rules, then the resource is committed to work correctly. The client-centric consistency insures that a client does not to come front with an inconsistency while having access to a resource. However, if multiple clients gain access to the same resource simultaneously, then the consistency is not guaranteed \cite{Steen2009}. Consistency models in the distributed systems are executed with different methods on variant machines, accordingly they use different methods of consistency i.e., data-centric, client-centric, and combination of the both \cite{Li2012a, mahajan2011depot}. For example, the data consistency model can be applied on the distributed shared-memory \cite{Lamport1978a}. \cite{Mosberger1993a, Lamport1979b, Keleher1992a, Dubois1986, Gharachorloo1990, bershad1991shared, ahamad1990causal}, or it might be executed on the shared data-center or the distributed database \cite{Papadimitriou1979c}, or the consistency model on the read and write operation of the clients or the data stored in the cache memory \cite{Alonso1990}, consistency can even be presented to create a session and its relation with the other sessions \cite{Terry1994c}. Message send and receive \cite{Liu2014} also needs the consistency to be performed. In spite of diverse applications in the scope of the consistency of the distributed systems, researchers seek to discriminate the performance of the different types of consistency. Little research has been done on reviewing and the promotion of different consistency types. Furthermore, with respect to the concepts of consistency, they have analyzed all of the data-centric and client-centric consistency from two perspectives. The introduced consistency models are analyzed based on the consistency model presentation time with respect to the strong to weak consistency and the capability of the consistency. Finally, they have mapped the definitions of consistency  on variant specific systems \cite{Lee2015}. By analyzing the data-centric and client-centric models or a combination of both based their usage, they are divided into the three categories of architecture based, distributed database and the distributed systems \cite{aguilera2016many}. The concepts of the consistency models introduced in this research are the traditional consistency models with the emerge of the distributed systems.
	
	\section{Contribution}\label{sec.Contri}
	In this section, in order to show the models' behavior without the need for the implementation, we will present the mechanism of these models in mathematical terms. Then, we will introduce the novel consistency models based on the new requirements in the distributed computing systems. As mentioned before, the consistency models based on their application are divided into three categories. Our main focus in this research is to define different consistency models which are used in the distributed systems.
	
	Different traditional consistency models are proposed in Fig.\ref{fig.timeLine}. In this new categorization, with respect the focus of this research on distributed systems, our goal is to introduce
	various types of consistency models such as data-centric, client-centric, hybrid, novel and extended followed by their implementation and evaluation environments between 1979 to 2018.

	Also with respect to this categorization we have shown the cruciality of consistency in the distributed systems. With respect to Fig.\ref{fig.timeLine}, the traditional consistency model has been proposed from 1979 to early 2000. By time and with the turn of the century (year 2000) and with the extense and improvement of the distributed systems, consistency models become mandatory. From the year 2000 to 2006 researchers proposed different models of data-centric consistency like the linearizability \cite{Goodson2002} and the extended consistencies such as the timed causality consistency \cite{Torres-Rojas2005} and a session and causal consistency \cite{brzezinski2004session} as a hybrid consistency have been studied. Researchers have proposed novel consistency models with respect to the specific needs. One the newly emerged consistency models is the fork consistency \cite{Li2004}. However, by time and with changes in essentialities, distributed systems gained more tendency in using the hybrid consistency models. Whilst, some others have proposed novel consistency models such as view consistency (VC) \cite{Lee2007} or the RedBlue consistency \cite{Li2012a}. Amongst which, consistencies like the fork consistency \cite{Feldman2010} and causal + consistency (CC+) \cite{Lloyd2011} are also introduced as extended consistency models. But, at this stage the need of the great distributed systems to the eventual consistency to reach appropriate accessibility has been observed. During the years 2012 through 2018 the researched have proposed a variety of different consistency models. For instance, the data-centric consistencies are the causality consistency \cite{Liu2014}. \cite{Lady2014, HosseinHaeri2016}, linear constancy \cite{Lee2015, Kleppmann2015}, and weak consistency, eventual consistency model \cite{Vukolic2016}, hybrid consistencies \cite{Brandenburger2015}, and some other novel consistencies such as the VFC3 \cite{Esteves2012}. What is shown in Fig.\ref{fig.timeLine}, is the growing demand of the distributed systems to the eventual consistency from 2006 to this date. As it can be seen, the distributed storage systems are proposed in order to facilitate the development and to have more precise evaluations on the consistency models. These storage systems have had lots of growth in the recent years. During 2006 through 2012 the environments such as Sanfonia \cite{aguilera2007sinfonia}, COPS \cite{Lloyd2011}, G-Store \cite{das2010g}, pahoehoe \cite{Anderson2010}, Gemmi \cite{Li2012a}, ALPS \cite{Lloyd2011}, and from 2012 to this date, the environments like Eiger \cite{Lloyd2013}, Orbe \cite{Du2013b}, FaRM\cite{dragojevic2015no}, Pileus \cite{Terry2013a}, RamCloud \cite{Lee2015}, H-Store \cite{Lee2015} are the novel storage systems that are proposed by the researchers.
	
	In this research, different types of consistency from traditional to novel consistency models are analyzed. We have described the performance of the traditional consistency systems in mathematical terms. The performance of different models used based on its application in the distributed systems environments is determined by the traditional consistency models section, then the novel consistency models which are proposed in the distributed systems discipline are defined in the novel consistency models. 

What is expected in this research is to review the challenges and issues in consistency on distributed systems.

	\section{Traditional consistency model}\label{sec.ConvConsis}
	
	As discussed before, the consistency model is one of the most important issues in the in the design of the storage systems in large scales \cite{Lee2015}. Generally, the data consistency could be divided into two categories of the data-centric and client-centric. However, today the researchers seek for the ways to present a hybrid of the consistency models with respect to the requirements of the applications. One of the main goals of this research is to introduce some of the consistency models which in contrast to the hybrid models not only are capable of ensuring the consistency in distributed storage systems, but also are able to cover those needs of the applications which are usually answered by the hybrid consistency models and not to mention have less weak points in comparison with the other models. The traditional consistency models are categorized in a specific class. The behavior of the processes is conveyed by the consistency model based on the model type and with the data items in the shared-memory through mathematical terms. Before we analyze and introduce the other types of the consistency, let us have a brief but complete introduction on the notations used in this paper:
	
	\begin{table*}[t]
		\caption{Symbols, notations and the keywords.}
		\label{tab.01}       
		\begin{tabular}{ll}
			Symbol & Description \\ 
			\hline 
			$x$ & read or write operation on replica $ x $. \\ 
			$W(x)$ & write operation on replica $ x $\\ 
			$R(x)$ & read operation on replica $ x $\\ 
			$A$ & the value of the data written on or read from the replica\\ 
			$B$ & the value of the data written on or read from the replica\\ 
			$O$ &  any read or write operation.\\
			$S$ &  the server that contains replica X\\ 
			$C$ &  the client that executes the operation on the replica\\ 
			$OW$ &  the primary set of operations on the shared data\\ 
			$OS$ &  the set of operations which are conducted by the server on the replica\\ 
			$OC$ &  the set of operations which are conducted by the client on the replica\\
			$P$ &  the process which executes the read or write operations on the replica\\ 
			$T_P$ &  the absolute global clock or the physical clock on the servers\\ 
			$T_S$ &  the logicalal clock \cite{Lamport1979b}, based on the occurrence by the process, server/client\\ 
			$\epsilon$ &  the elapsed time, immediately after the precedent operation $ (\epsilon<\delta<\gamma) $\\ 
			$\delta$ &  the elapsed time after the precedent operation $ (\epsilon<\delta<\gamma) $\\
			$\gamma$ &  the long elapsed time after the precedent operation $ (\epsilon<\delta<\gamma) $\\
			$LS$ &  the process which posses this lock can perform the read or write operations on the replica $ x $ \\
			$L$ &  the process which posses this lock can perform the read or write operations on the replica $ x $ \\
			$ Acquire(x, l) $ & the acquirement of the lock $ L $ for the execution of the operation on the replica $ x $\\ 
			$ Release(x, l) $&  the release of the existing lock $ L $ on the replica $ x $ after the termination of the operation on \\
			& the specified replica\\ 
			$ L_x $ &  the lock for the execution of the operation on the replica $ x $\\ 
			$ L_y $ & the lock for the execution of the operation on the replica $ y $\\ 
			$ e1 \xrightarrow{s_i} e2 $ & operation $ e1 $ is initially performed on the server $ S_i $  and then the operation $ e2 $  \\ & is performed on the same sever\\ 
			$ e1 \pdp e2 $ &  the process $ P_i $ executes  operation $ e1 $ initially and then executes operation $ e2 $\\ 
			$ a \Rightarrow b $ &  if the set of operations $ a $ occurs, then the set of operations $ b $ occurs\\ 
			$ a \longmapsto b $ &  if any operation $ a $ occurs, then operation b must occur\\ 
			\hline
		\end{tabular}
	\end{table*}

	\subsection{Data-centric consistency model}\label{sec.DataCentric}
	
	\subsubsection{Strict consistency model}
	
	Strict consistency is the strongest consistency model which requires permanent global synchronization. This synchronization is done by using an absolute global time. The creation of this synchronization by the physical time among the servers is to some extend impossible \cite{Steen2009}. In other words, the replicas must be synchronized globally and constantly. This model is so costly, while the system does not need to be always synchronized globally.  However, in the distributed systems this straightforwardness imposes lots of expenses. With respect to the behavior of consistency model on the replicas, we can express the behavior of this system, using \ref{rule.01}.
	
	\begin{matriz} \label{rule.01}
		\begin{split}
				\underset{w(x)a \in OW \cup OS_i, r(x) a \in OW \cup OS_i}{\forall}\\
			\begin{bmatrix}
				w(x, T_p) a \xrightarrow{S_i} r(x, T_p + \epsilon) a \bigwedge\\
				\underset{w(x)b\in OW \cup OS_i}{\nexists}  \\
				\begin{pmatrix}
					a \neq b, \delta<\gamma, w(x, T_p)a \\
					\xrightarrow{S_i} w(x, T_p)a \\
					\xrightarrow{S_i} w(x, T_p + \delta)b \\
					\xrightarrow{S_i} r(x, T_p + \gamma)a
				\end{pmatrix}
			\end{bmatrix}
		\end{split}
	\end{matriz}
	
	with respect to \ref{rule.01}, if the write operation $ a $, done in time $ T_p $ on the server $ S_i $ is terminated by a process, it could also be executed by the read operation $ a $ in time $ T_p + \epsilon $ by the same or other process, in a time immediately after the write operation $ a $ is performed. If a new value of $ b $ is occurred by a process on the server $ |S_i $ in time $ \delta $ after the write operation of $ a $, then the read operation of $ a $ in time $ \gamma $ after the write operation of $ b $ is performed, shows the strict inconsistency. In order to get more familiarized with the interaction of this model concept and process with data items in the shared-memory, the strict consistency model is shown in Fig.\ref{fig.Behavior}. When the process $ P_1 $ executes the value $ b $ by writing on the servers in time $ \delta $ (a time after each reading or writing process of the value $ a $ from the shared-memory), but in time $ t_8 $ by having access to the server $ s_1 $ or in time $ t_9 $ by having access to each of the servers $ s_1 $ and $ s_2 $, by execution of the read operation, again the stale value of $ a $ is placed in the value $ b $'s place to be read by the operation $ p_3 $. In this case, the deficiency in the shared-memory is uncovered by the strict consistency. 
	

	\subsubsection{Sequential consistency model}
	
	 Another variant of the consistency models is the sequential consistency. This model was firstly defined by Lamport in 1979 \cite{Lamport1979b}. In the discipline of the shared memory for the multi-processor systems. This model is a simpler variant of the strict consistency model. In the sequential consistency model the need to the absolute global time and the full-time relation of the replica are not the same as the strict consistency model. In this model, the write operation is viewed between the processes with an equal order, however the read operation done by the other processes is not observable \cite{Steen2009}. The concurrent execution of multiple write operations without having the causality relation by different processes is observed with different orders \cite{A2015}. Thus, this model guarantees that the values are read by the clients \cite{Torres-Rojas2005, Bermbach2013}. The malicious server, is a server in the distributed environment in which the sequence of the operations is not preserved. In the systems where it is probable for the malicious server to be existent, the execution of the operation is by a healthy service provider is acceptable \cite{Brandenburger2015, Li2004, mahajan2011depot, Brandenburger2017, Kong2005, Feldman2010, Shraer2010, Williams2009a}, when its proportional order is preserved for each client, however it cannot guarantee the truth and the complete execution of the whole operation by the clients \cite{Cachin2007}. With respect to the CAP theorem \cite{Brewer2012}, the restrictions in the distributed systems (consistency, availability, partition tolerance) have been described in a way that the sequential consistency could be provide in the system \cite{belaramani2006practi}. The sequential consistency have shown better performance on the history of the operations in the conflict serializability \cite{Brutschyc}. Finally, the confirmation of the sequential consistency \cite{Steen2009} is considered by the properties like  the atomicity of NP-complete \cite{Golab2011}. Based on the behavior and the performance of the operations on the replica in the sequential consistency model, the following equation could be expressed:
	 
	 \begin{equation}\label{eq.01}
	 	\forall_x (\forall s_i w(x)a \xrightarrow{S_i} w(x))b \vee \forall s_i w(x)b \xrightarrow{s_i} w(x)a) 
	 \end{equation}
	 
	 With respect to eq. \ref{eq.01}, the time parameter does not affect the behavior of the process, however, the sequence of the operations on the server has a considerable effect on the shared-memory. We have expressed the performance of the processes on the data items in the shared-memory with the sequential consistency in terms of \ref{rule.02}.
	 
	 \begin{matriz} \label{rule.02}
	 	\begin{split}
	 		\underset{w(x)a \in OW \cup OS_i, r(x) a \in OW \cup OS_i}{\forall} \\
	 		\begin{bmatrix}
	 			w(x)a \xrightarrow{S_i} w(x)b\\ 
	 			\wedge \underset{r(x)b \in OW \cup OS_i}{\forall} \\
	 			 \begin{pmatrix} 
	 				a \neq b, w(x)a \xrightarrow{S_i} w(x)b\\
	 				\xrightarrow{S_i} r(x)a \xrightarrow{S_i} r(x)b
	 			\end{pmatrix}\\ \wedge
 			\underset{ r(x)b \in OW \cup OS_i}{\nexists}\\
 			\begin{pmatrix}
 				a \neq b, w(x)a \xrightarrow{S_i} w(x)b\\
 				 \xrightarrow{ S_i} r(x)b
\xrightarrow{S_i} r(x)a 			\end{pmatrix}
	 		\end{bmatrix}
	 	\end{split}
	 \end{matriz}
 
 If the write operation of the values $ a $ and $ b $ on the server $ S_i $ based on \ref{rule.02} be $ a $ and then $ b $. Also, the read operation of these values by accessing the server $ S_i $ be $ a $ and $ b $ respectively. If the whole processes do not read $ a $ and then $ b $ consequently, then they would not have the same vision of the whole operation and values stored in the shared database. Therefore, the sequence of observation is different and the inconsistencies are revealed. The interaction of the processes with replica in the shared database with the sequential consistency model is shown in Fig.\ref{fig.Behavior}

A process like $ P_5 $ causes inconsistencies, as it has a different view in reading from the shared-memory. In such a way that this process in contrast with the other processes (e.g., $ P_3 $ and $ P_4 $), with the execution of the reading process, firstly read the stale value of $ a $, and then reads the value of $ b $. The execution of this process by the process $ P_5 $ causes inconsistencies in the shared database with sequential consistency.

\subsubsection{Linearizability model}

One other type of the data-centric consistency models is the linearizability, which is also known as the strong consistency model. This model is considerably better than the sequential consistency which is introduced in 1990 by Morris Herlihy and Wing in 1990 \cite{Herlihy1990}.  This model also needs a global synchronization clock. As this clock is not as trust worthy, it is replaced by logical clock in this model \cite{Lamport1979b}, which is called the global logical time. The behavior of this model is like the sequential consistency model. However, the order of the operations is set by the whole processes based on their time of occurrence. In this case, by putting a limitation on the sequential consistency model, e.g., the event time, this model changes to the linearizability, which has a significant effect on the write operation on the shared-memory.  Therefore, the whole processes based on this consistency have a solid view of the operations. This behavior is like the serial and the sequential behavior by a server \cite{Saito2005, Li2012a, Bernstein2013a, Bailis2012e}. In order to state the sequential operation and find a solution to avoid the controversial operations, the linearizability model is used instead of the sequential consistency \cite{Bouajjani2014}. The consistent services require the linearizability in order to make sure of the complete execution of the operations on the untrustworthy systems of the Replicated State Machine (RSM) \cite{Poke2015a}. This model uses the Wait-Freedom approach in case the system is untrustworthy \cite{Cachin2009}. Linearizability is a criterion to analyze the accuracy of the operations and the degree of reliance the clients can have on the distributed storage systems when they are faced with an untrustworthy server \cite{Bernstein2013a, Cachin2009}. In order to guarantee the linearizability, the secure network protocols, the automatic repetition of the failed operation, and the two-phased commit protocols are used \cite{Lee2015}. In spite of the fair-loss, the linearizability, in a system with network partition is an acceptable algorithm as it might wait for the partition to get better which never does happen. Therefore, the system model with unlimited partition is acceptable \cite{Kleppmann2015}. In this case, when the system is faced with network partition the system would have an acceptable access limit. This consistency model requires that each of the read or write operations to be done in an interval between the operation execution request and response \cite{Vukolic2016}. Finally, the verification of the linearizability is an NP-Complete problem \cite{Golab2011}. What we stated up to now about the behavior of the processes in the linearizability model is according to \ref{rule.03}, trough which the whole behavior and the manners of the processes are considered thoroughly.

\begin{matriz} \label{rule.03}
	\begin{split}
		\underset{w(x)a \in OW \cup OS_i,r(x) a \in OW \cup OS_i}{\forall}\\
		\begin{bmatrix}
			w(x, T_S)a \xrightarrow{S_i} w(x, T_S + \epsilon)b\\ \longmapsto \begin{pmatrix}
				r(x, T_S + \delta)a \\
				\xrightarrow{S_i} r(x, T_S + \gamma)b
			\end{pmatrix}\\ 
			\wedge 
			\underset{r(x)b \in OW \cup OS_i}{\nexists} \\
			\begin{pmatrix}
				a \neq b, \delta < \gamma, w(x, T_S)a\\
				 \xrightarrow{S_i} w(x, T_S + \epsilon)b\\
			 \xrightarrow{S_i} r(x, T_S + \delta)b \\ \xrightarrow{S_i} r(x, T_S + \gamma)a		\end{pmatrix}
		\end{bmatrix}
	\end{split}
\end{matriz}

\begin{figure}
 	\includegraphics[width=\columnwidth]{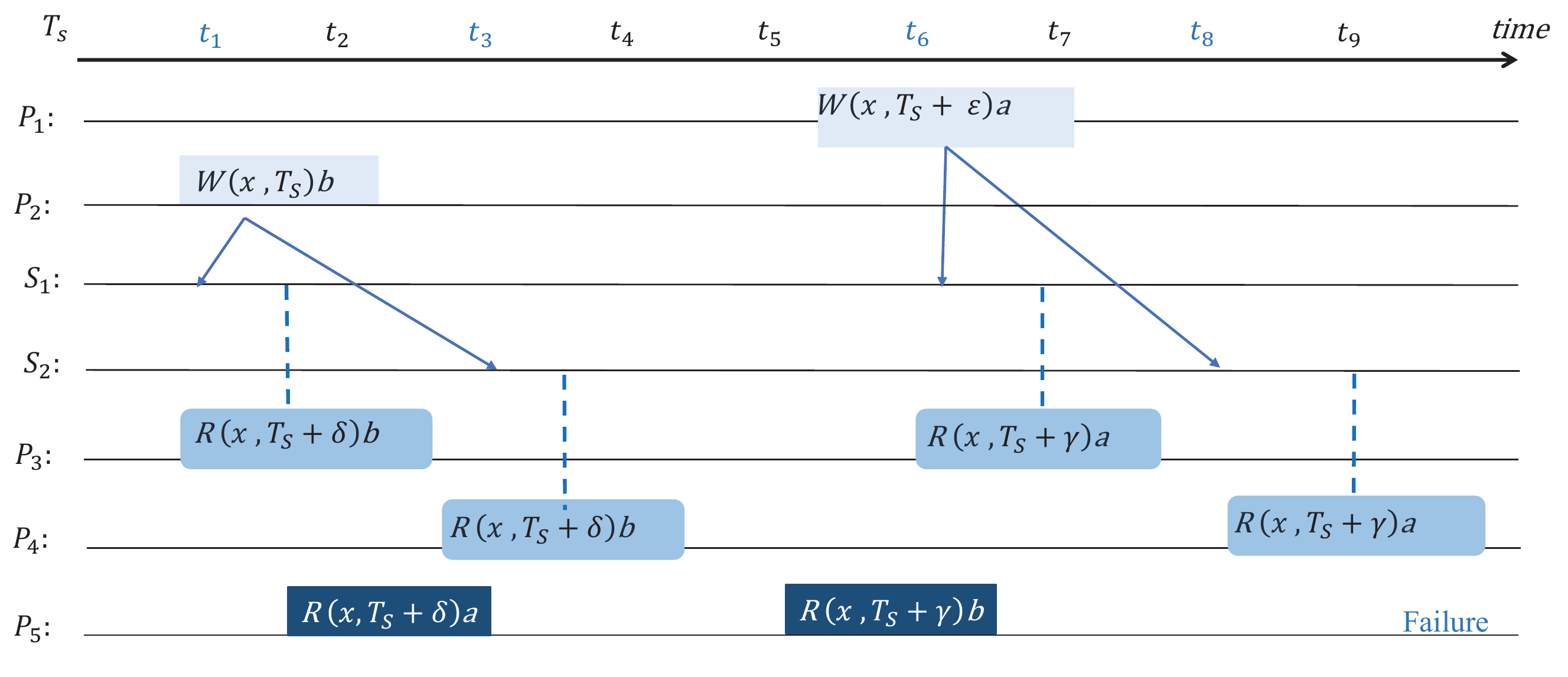}\\
 	\centering{
 		\caption{The behavior of the processes on the data-items based on the linearizability model.}
 		\label{fig.BehaviorLinear}
 	}
 \end{figure}

 The order and priority of the execution is equal for the whole processes according to \ref{rule.03}, and is based on their logical event time. When the write operation of b is executed in time $ \epsilon $, after the write operation of $ a $ on the shared-memory in the sever $ S_i $, then it is expected that after having access to the data storage, each process by the execution of the read process read the value a in time $ \delta $, and subsequently, the value b in time $ \gamma $ from the server. If trough the update process of the data storage, the new value of b is written in time $ \epsilon $, and then in time $ \delta $, with the execution of the read operation, the server return the same new value (i.e., the value of $ b $), then the controversy is emerged in the linearizability. To have a deep understanding of this consistency model, its operating process is illustrated in Fig.\ref{fig.BehaviorLinear}.
 
 Based on what is depicted in Fig.\ref{fig.BehaviorLinear}, if the process $ P_s $ act contradictory to the priority of the write process by the processes $ P_1 $ and $ P_2 $, firstly by the execution of the read operation in time $ \delta $ return the value $ b $, then the data storage has violated the linearizability model.
 
\subsubsection{Causal consistency model}
 
This model has been first proposed for the shared distributed system \cite{ahamad1990causal} and is weaker than the sequential consistency model which was proposed in 1987 based on the Happened - Before by Lamport for the distributed systems \cite{Lamport1978a}. This consistency model discriminates between the events which have cause and consequence relationship and those which does not have. In case the read operation is the result of multiple write operations on the replica, then the read operation does not execute until the write operation has been terminated \cite{Saito2005}. This model guarantees that a client does not read two related write operations in a wrong order and in case the client has read the latest value, then it does not read the stale value \cite{Shraer2010}. If the process updates a replica and the proceeding process return the updated value, this relationship between the two processes is provided by the causal consistency \cite{Vogels2008}. This model could be defined as a combination of the causal arbitration and causal visibility \cite{Burckhardt2014}, which in spite of the partitioning of the network has satisfaction of the availability \cite{Burckhardt2014, Brewer2012}. To implement this consistency model, we need the logical time to record each event \cite{Lamport1979b}. As the global time is not considered in this model, then the causal consistency model is not solely capable of providing the convergence. To solve this problem, the time factor has been added to this model which is called the timed causal consistency \cite{Torres-Rojas2005a}. In all models Complete Replication and Propagation protocol is used \cite{hsu2018causal} Specifically the full replication simplifies the causal consistency on data-items, operations and replicas. By converging the conflict operations, the causal consistency is called the Causal + Consistency (CC+). In geo-replicated systems or online applications like the social networks where the operations must be executed completely and with low latency, the CC+ insures the clients that they see the cause and effect relation correctly, without confliction and with steady process in the storage system \cite{Lloyd2011}. This relations between the clients' operations, the data or the keys stored in the shared memory, the sessions which the clients use and the log files existing in the replica have causality. The operations in this model are expressed as in eq. \ref{eq.02}

\begin{subequations}\label{eq.02}
	
	\begin{equation}
	O1 = w(x)a \wedge O2=r(x)a
	\end{equation}
	    
	\begin{equation}
	O1 = w(x)a \wedge O2=w(x)b
	\end{equation}
	
	\begin{equation}
	O1 = r(x)a \wedge O2=w(x)b
	\end{equation}
\end{subequations}

Eq. \ref{eq.02} shows the concentration of this model on the conflicting operations. The goal of this model is to express the type of relation between these operations. This consistency model is determined based on the type of the cause and effect and the priority of the relation. In case the relations according to eq. \ref{eq.03} have dependency to each other, then the whole processes, with respect to the event time and the dependency of the operations \cite{Liu2014, Zawirski2015} will execute the operations with a united perception.

\begin{equation}\label{eq.03}
	\underset{P_i}{\exists} O1 \xrightarrow{P_i} O2
\end{equation}

With respect to eq. \ref{eq.03}, the whole processes initially observe the first operation and then the second. In case the operation $ o $  occur between $ o1 $ and $ o2 $ and play the role of an intervener between the two operations, then the following equation stands \cite{brzezinski2004session}:

\begin{equation}\label{eq.04}
	\underset{o\in O}{\exists} (o1 \rightsquigarrow o \wedge o \rightsquigarrow o2)
\end{equation}

In eq. \ref{eq.04}, the operation $ o $ indicates the intervening operation which establishes the cause and effect relation between the two operations $ o1 $ and $ o2 $. In this case the concurrent operation could be expressed as in eq. \ref{eq.05}.

\begin{equation}\label{eq.05}
	\underset{P_i}{\nexists} O1 \xrightarrow{P_i}O2 \vee O2 \xrightarrow{P_i}O1 \Rightarrow O1 \parallel O2
\end{equation}

However, if the operations are independent to each other, then the whole processes could behold the operations with their own vision. If two processes simultaneously and automatically write to different data-items or read a shared data-item, then the processes will not have a cause and consequence relation and the operation is called concurrent \cite{Tanenbaum2007}. If the operation $ o1 $ and then $ o2 $ are executed by each process $ P_i $ or vice versa, then this kind of behavior implies that there are no cause and effect relations between the operations and therefore they could run simultaneously. Thus the causal consistency model could be expressed by the \ref{rule.04} which shows the behavior and the performance of this model on the shared-memory.

\begin{matriz}\label{rule.04}
	\underset{S_i}{\forall} o1, o2 \underset{Ow \cup O_{S_{i}}}{\forall} (o1 \rightsquigarrow o2 \Rightarrow o1 \xrightarrow{S_i}o2)
\end{matriz}

\begin{figure}
	\includegraphics[width=\columnwidth]{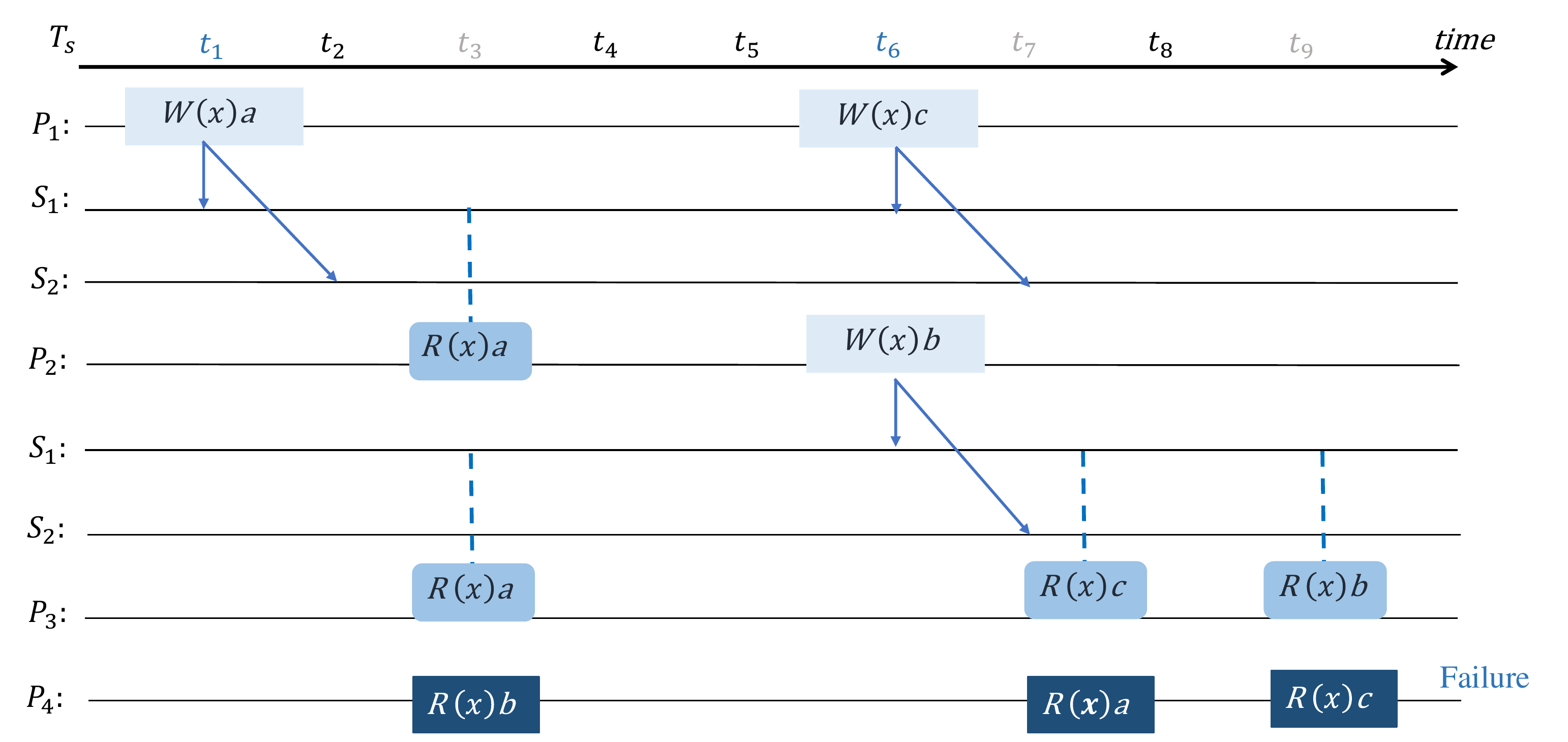}\\
	\centering{
		\caption{The behavior of the processes on the data-items based on the causal consistency.}
		\label{fig.BehaviorCausal}
	}
\end{figure}

This rule expresses that in case the execution of the operation $ o1 $ is the cause of the execution of the operation $ o2 $ on the replica in the sever $ S_i $, then the other processes must first observe the operation $ o1 $ and then the operation $ o2 $ on their own server \cite{brzezinski2004session}. 
An example of the interaction of the processes with the data-items stored in the shared-memory under the coverage of the causal consistency is shown in Fig.\ref{fig.BehaviorCausal}.

Fig.\ref{fig.BehaviorCausal} illustrates two different operations based on the cause and effect relations between the operations by the causal consistency model. First, the event of simultaneous write operation of $ b $ and $ c $ by two different processes in time $ t_6 $ where there are no cause and effect relations between them. In this case, these two operations could be perceived differently by the other processes. However, the main point is the relation between the read operation of the value $ a $ by the processes $ P_2 $ and then is the start of the write operation of the value $ b $ on the server $ S_i $. Thus, the whole processes need to have access to a server to read the value $ b $, on which the value of $ a $ is already written and then value of $ b $ is written on the same sever in order to finally read the value $ b $. Otherwise, there would be a violation in data storage in terms of the causal consistency.

\subsubsection{The first-in, first-out consistency model}

The first-in, first-out (FIFO), also known as the pipelined random access memory (PRAM) is another type the data-centric consistency models. This model was proposed by Lipton and Sandberg in 1988 \cite{Mosberger1993a}, in which the sequence of the write operations of a process is desired when multiple operations start to write on the data storage by multiple processes, the other processes see the operations with equal sequences which are done by a process and might see the operation of the other processes differently in terms of sequence. One of the applications of this model is to give privileges to the pipelined write operations \cite{Mosberger1993a}. The healthy servers send messages to the clients and this message sending is done with the order and sequence of the FIFO consistency model. Each client, communicating with the server, could receive the messages through the asynchronous secure channel according to the FIFO consistency model \cite{Cachin2011, Brandenburger2015, Cachin2009}. Therefore, the client receives the message by reliance on the server; however, the malicious server rearranges the messages which causes a delay in sending the message and as a result its deletion. This time delay and deletion testify the existence of a malicious server \cite{Lady2014}. according to eq. \ref{eq.02}, \ref{rule.05} could be written as the behavior of the FIFO consistency model:

\begin{matriz}\label{rule.05}
	 \underset{\underset{P_i}{\exists} o1, o2 \in Ow \cup O_{S_{i}}}{\forall} (o1 \xrightarrow{P_i} o2 \Rightarrow o1 \xrightarrow{S_i} o2)
\end{matriz}

\begin{figure}
	\includegraphics[width=\columnwidth]{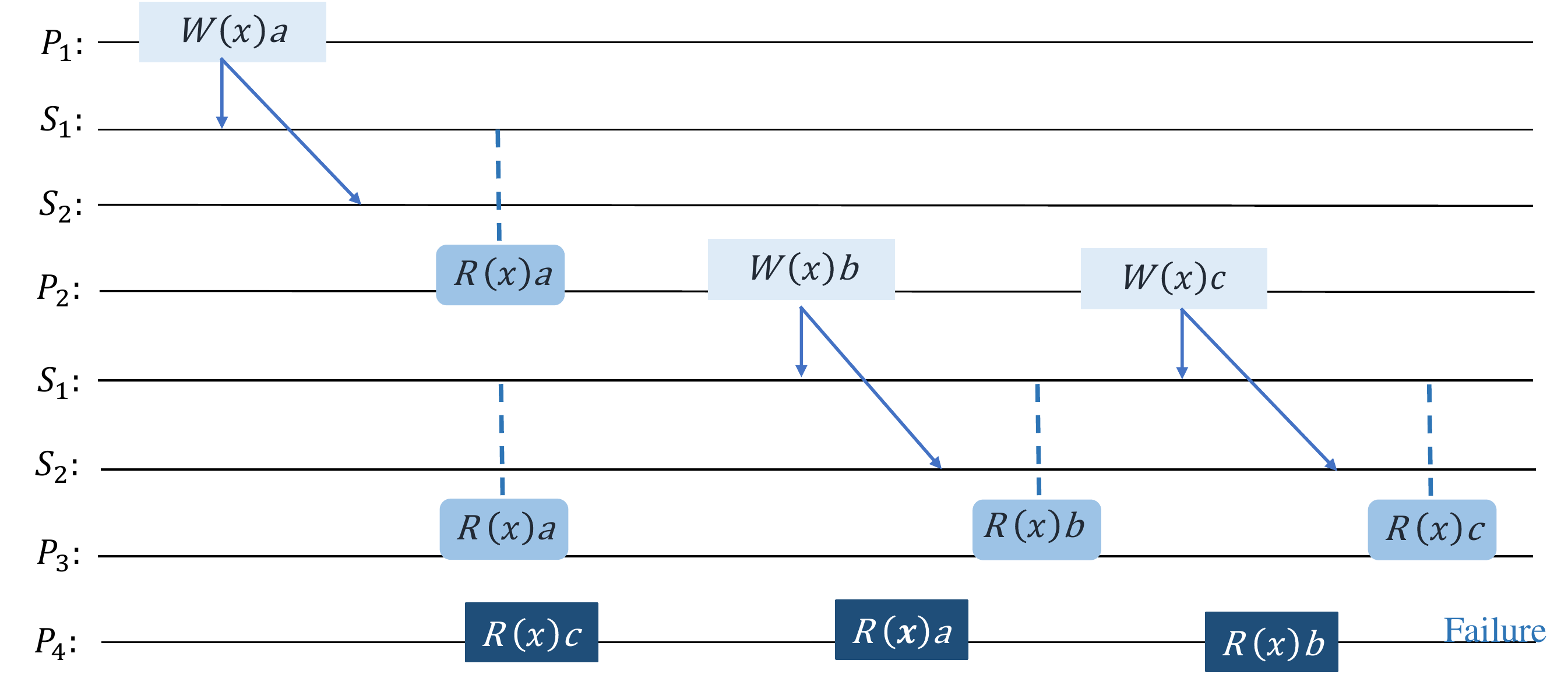}\\
	\centering{
		\caption{The behavior of the processes on the data-items based on the FIFO consistency.}
		\label{fig.BehaviorFIFO}
	}
\end{figure}

If the operations $ o1 $ and $ o2 $, according to \ref{rule.05} are executed by a process and with respect to the priority first the operation $ o1 $ and then the operation $ o2 $ are preformed, on the side of the server $ S_i $, first the operation $ o1 $ and then the operation $ o2 $ are observed. The interaction of the processes with the data-items in the shared-memory using the FIFO consistency model is depicted in Fig.\ref{fig.BehaviorFIFO}.

One of the important points in this consistency model is its independence of time in prioritizing the operations. From what is illustrated in Fig.\ref{fig.BehaviorFIFO}, the sequence of the writing of the values $ b $ and $ c $ by the process $ P_2 $ is to the write operation of $ b $ at first and then the operation$ c $ on the server $ S_i $. The goal is to observe this sequence between the whole processes. In case like the process $ P_4 $ this sequence is different with the observation of the other processes, then there would be violations in the data storage in terms of the FIFO consistency.

\subsubsection{Weak consistency}

This consistency model is another variant of the data-centric models which was proposed in 1986 by Micheal Dubois \cite{Dubois1986}. The weakness of this consistency model is the reason behind this difference. This avoids the global synchronization. To do so, it defines a synchronization variable which plays the role of token. The process which possesses the this token could preform the read or write operations on the shared resource. In this model, the accessibility to the resource is done by the sequential consistency within the synchronization variable. If the process does not possess this token neither of the read or write operations are privileged on the shared resource \cite{Mosberger1993a, Dubois1986}. This model is established under the following conditions \cite{Dubois1986, bouge2013managing, A2015}:

\begin{itemize}
	\item Accessibility to the synchronization variable by the whole processes (i.e. nodes or processors) is not done with an accordant sequence.
	
	\item The other accessibilities might be observed differently by the other processes.
	
	\item A set of both read and write operations during the different synchronization operations is consistent in each process.
	
\end{itemize}

In this consistency model if multiple processes want to merely preform the read operation form the shared memory, then all of them can have the synchronization variable at their service. However, if a process want to preform the write operation on the shared memory, then until the termination of the write operation by the process, the other processes cannot have the synchronization variable up until the termination of the write operation. The behavior of the process when they are served by the synchronization variable is expressed by eq. \ref{eq.06}:

\begin{equation}\label{eq.06}
	\begin{split}
	\forall_x(\forall_{S_i} w(x, LS)a \xrightarrow{S_i}  r(x, LS)a \\ \longmapsto \forall_{S_i} w(x, LS)a \xrightarrow{S_i} O(x, LS)b)
	\end{split}
\end{equation}

The characteristic of the eq. \ref{eq.06} is based on the synchronization variable $ LS $. Eq. \ref{eq.06} shows that in order to execute any kind of operation $ O(s.LS)b $ either of read or write on the replica the possession of the synchronization variable is necessary. This variable is attainable only after the write operation is terminated by the process and the synchronization variable is free. In this model the local replica updates the other replicas after the termination of the final changes. What makes this consistency model to be at odds with the other types of consistency is in its toleration of the violence in consistency in the intervals between each two updates. By considering the receiving medium of the synchronization variable by the processes and the behavior of this consistency model \ref{rule.06} can be expressed as:

\begin{matriz} \label{rule.06}
	\begin{split}
		\underset{w(x)a \in OW \cup OS_i, w(x) a \in OW \cup OS_i}{\forall} \\
		\begin{bmatrix}
			\begin{pmatrix}
				w(x, LS)a \xrightarrow{S_i} w(x, LS)b\\
				\Rightarrow \forall_{r(x)b \in  ow \cup os_i} (r(x, LS)a \xrightarrow{S_i} r(x, LS)b)
			\end{pmatrix}\\
			\wedge 
			\underset{ r(x)b \in OW \cup OS_i}{\nexists}\\
			\begin{pmatrix}
				a \neq b, w(x, LS)a \xrightarrow{S_i} w(x, LS)b\\
				\xrightarrow{S_i} r(x, LS)a \xrightarrow{S_i} r(x)b
			 	\end{pmatrix}
		\end{bmatrix}
	\end{split}
\end{matriz}

\begin{figure}
	\includegraphics[width=\columnwidth]{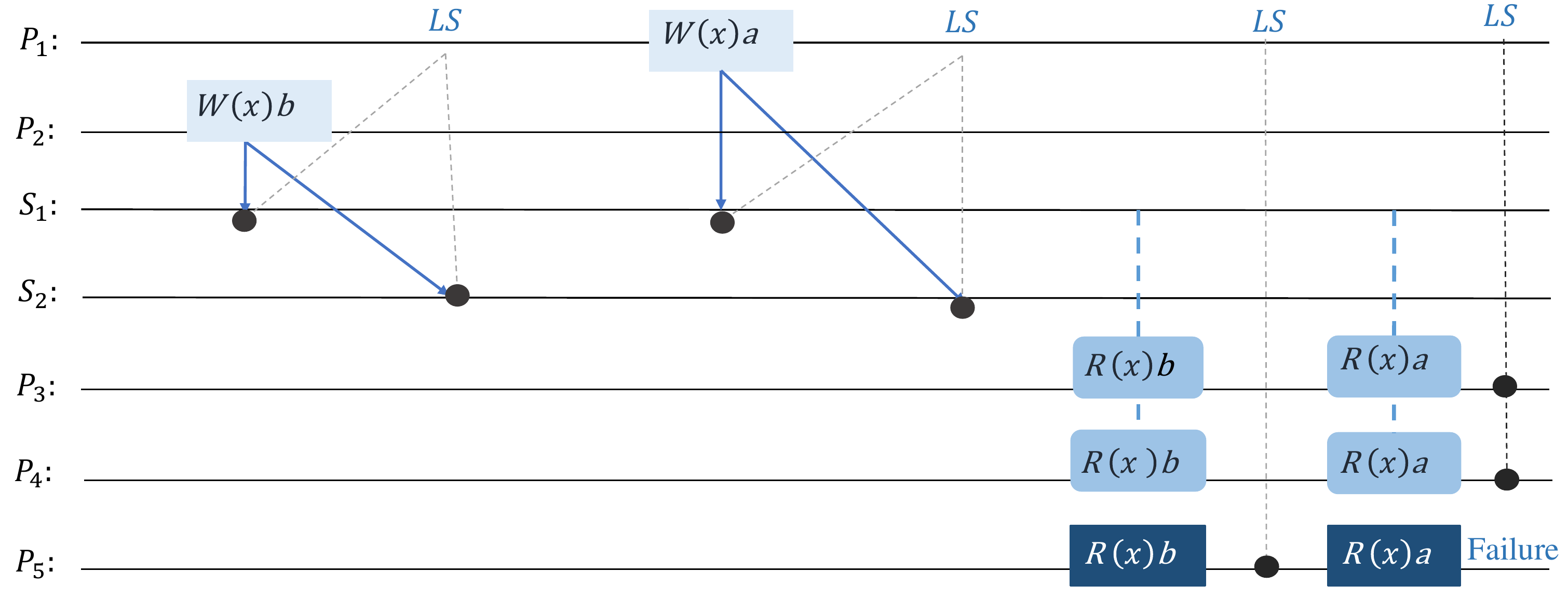}\\
	\centering{
		\caption{The behavior of the processes on the data-items based on the weak consistency.}
		\label{fig.BehaviorWeak}
	}
\end{figure}

Using the weak consistency model, \ref{rule.06} expresses that, in order to preform any operation on the memory, the violation occurs in data storage if process has not received the synchronization variable. When the process asks the server $ S_i $ to read the value of $ b $ without the synchronization variable, then the returned value might not be valid. Fig.\ref{fig.BehaviorWeak} depictss the behavior of this consistency model.

The red points in Fig.\ref{fig.BehaviorWeak} indicate the release of the synchronization variable after the termination of the operation on the replica existing in server $ S_i $. The point on which this figure concentrates is that the synchronization variable during the read operation is shared between multiple processes simultaneously. However while a process employs the synchronization variable for the write process, no other processes are able to have access to it. If a process like $ P_5 $ release the variable after the read operation, then if it wants to read $ a $ without receiving the variable, subsequently the read value of $ a $ is not valid and there would be violation in terms of the weak consistency in the shared-memory.

\subsubsection{Release consistency}

The release consistency is one of the data-centric weak models which was proposed in 1990 by K. Gharachorloo \cite{Gharachorloo1990}. This consistency model again uses the lock or the synchronization variable. In order to have access to the synchronization variable, two steps must be paved. In the first step it is necessary to ask for the lock to have access to the memory. Therefore, the process waits to have access to the replica stored in the memory. After receiving the lock which is symbolized by $ L $ in the table \ref{tab.01}, the lock is set on all the replica in the memory. In case the lock is not received by the process during the execution, then the obtained results are not valid. In the second step, the lock received by the process which asked for it is released and the updated values in the replica is sent to the other replicas in other servers in order to update their operations and values \cite{Mosberger1993a}. The problem emerging in the weak consistency is that in the access time to the synchronization variable, the distributed shared memory does not have any idea of the operation (i.e. read or write) on the replica. This problem has been solved by the release consistency. In this model the type of the operation is determined by receiving the lock and after its release. This model is not guaranteed for the geo-distributed systems with high availability \cite{Diack2013}. The coarse-graininess of the communications and the coherence of the systems with virtual shared-memory with error handling in sharing extensive communication by the release consistency model is one of the applications of this model \cite{iftode1996amd}. \ref{rule.07} states the behavior of the processes in the shared memory using the release consistency model. 

\begin{matriz} \label{rule.07}
	\begin{split}
		\underset{w(x)a \in OW \cup OS_i, w(x) b \in OW \cup OS_i}{\forall} \\
		\begin{bmatrix}
				w(x, l)a \xrightarrow{S_i} r(x, l)a\\
			\wedge
			\underset{r(x)a \vee r(x) \in OW \cup OS_i }{\nexists}\\
			\begin{pmatrix}
				a \neq b, w(x, l)a \xrightarrow{S_i} w(x, l)b\\
				\Rightarrow r(x, l)a \vee r(x)b
			\end{pmatrix}
		\end{bmatrix}
	\end{split}
\end{matriz}

If according to \ref{rule.07} a process take the synchronization variable, then its operation is valid. Therefore, by receiving the synchronization variable, the process takes its demanded replica and updates it. The read operation like the write operation needs to receive the synchronization variable and by holding it and the execution of the read operation it reads a valid value. Otherwise, without receiving the synchronization variable the written or read value is not valid. The conditions under which the release consistency could be executed are as follows:

\begin{itemize}
	\item The process needs to successfully put the lock on the shared-memory before preforming the read or write operations.
	
	\item Before releasing the lock on the memory, the read or write operation must be terminated by the process which holds the lock.
	
	\item The accessibility to the synchronization variable needs to be done using the first-in, first-out consistency model.
\end{itemize}

\begin{figure}
	\includegraphics[width=\columnwidth]{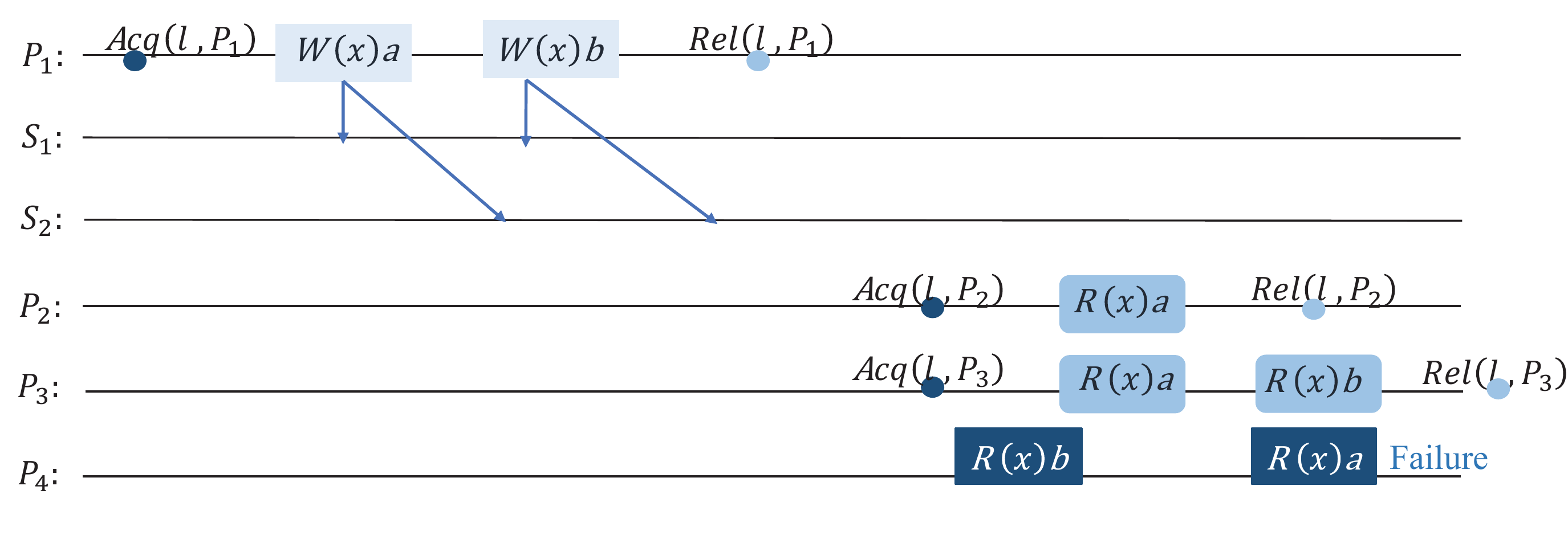}\\
	\centering{
		\caption{The behavior of the processes on the data-items based on the release consistency.}
		\label{fig.BehaviorRelease}
	}
\end{figure}

The behavior of the process in the shared-memory is sketched in Fig.\ref{fig.BehaviorRelease}.

In this model, the process which has received the lock is also determined and after the termination of the operation it releases the lock and the characteristics of the process are also eliminated. In case a process like $ P_4 $ does not receive the synchronization variable, then the memory faces with violations in terms of the release consistency.

\subsubsection{Lazy release consistency}

The challenge in the release consistency is that after the termination of the write operation and the release of the synchronization variable, it should propagate the entire changes occurred on the data in the replica to the other replicas in the memory. However, in case the update happens in all the replicas, there might be some replicas which do not need the update. Then with the propagation of the update the overhead increase and as a result the performance of this model would not be efficient. The lazy release consistency model is an extension to the release consistency which has been proposed in 1992 by $ P $. Keleher \cite{Keleher1992a} in order to improve the efficiency and optimality of the release consistency model. In this model the update occurs in the other replica only when it really needs to be updated. If necessary, it will send a message to the replicas in which the data has already been modified and updated. It is important to note that the timestamp is a great help in determining whether the data is obsolete or accordant to the latest update.

\subsubsection{Entry consistency}

The entry consistency model which is a variant of the weak consistency models has been proposed in 1991 by Bershad \cite{bershad1991shared}. This consistency model like the release consistency uses the lock with the difference that is considers a lock for each data-item. One of the problems in implementing the entry consistency is to determine the data for the synchronization variables. In this model, by receiving the lock the access to the data-item stored in the replica is provided, and any type of operation (i.e. read or write), is applicable to that data-item. After the termination of the write operation, like the release consistency model, the lock is released. However, in case of the read operation the release of the lock is not possible as the process might need the lock to execute some other operations on that data-item. In absence of the lock and the execution of the read operation from the memory, the read value would be invalid. This model is the weakest data-centric consistency model as in the interval when the lock is released, the consistency is not executed on the other replicas and the process holding the synchronization lock applies the consistency to the replica in the shared-memory \cite{Tanenbaum2007}. \ref{rule.08} states the behavior of the entry consistency model.

\begin{matriz} \label{rule.08}
	\begin{split}
		\underset{w(x)a \in OW \cup OS_i, r(x) a \in OW \cup OS_i}{\forall} \\
		\begin{bmatrix}
			w(x, lx)a \xrightarrow{S_i} rel(lx)\xrightarrow{S_i}r(x, lx)a\\
			\wedge
			\underset{r(x)a \in OW \cup OS_i }{\nexists}\\
			\begin{pmatrix}
				x(x, lx)a \xrightarrow{S_i}rel(lx) \rightarrow w(x, lx)a\\ \rightarrow(x, lx)a \xrightarrow{S_i}rel(lx) \xrightarrow{S_i}r(x, lx)a\\
				\wedge\\
				\nexists_{r(x)a \in OW \cup OS_i}\\ (w(x, lx)a \xrightarrow{S_i } rel(x) \xrightarrow{S_i}r(x)a)		
			\end{pmatrix}
		\end{bmatrix}
	\end{split}
\end{matriz}

\begin{figure}
	\includegraphics[width=\columnwidth]{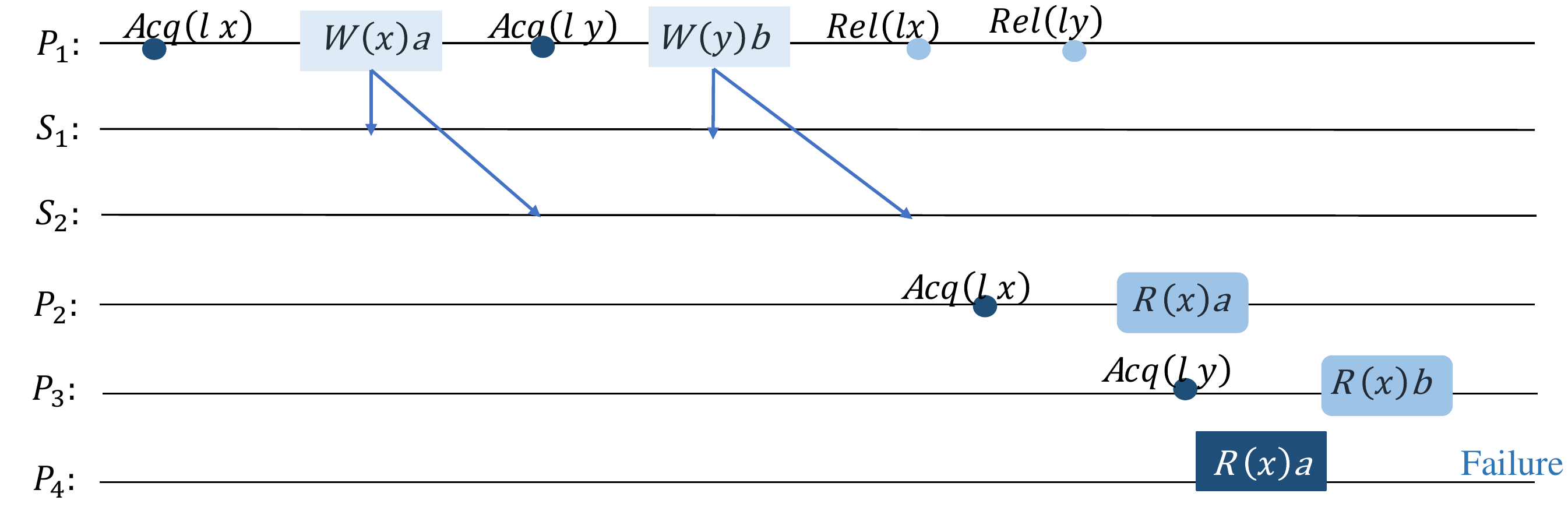}\\
	\centering{
		\caption{The behavior of the processes on the data-items based on the entry consistency.}
		\label{fig.BehaviorEntry}
	}
\end{figure}

The write operation in accordance with \ref{rule.08} suggests that at first the lock related to the particular data-item is received in the replica of the shared-memory and then it is released after the termination of the write operation. If a process for executing the read operation receives the lock of the data-item corresponding to that replica of the shared-memory, then with the read operation, it reads the valid value. When a process does not receive the lock corresponding to the desired data-item then the memory faces violations in terms of the entry consistency. In order to have a better understanding of how this model \ref{fig.BehaviorEntry} illustrates its behavior in interaction with the shared-memory.

As it is shown in Fig.\ref{fig.BehaviorEntry}, in case a process has accessibility to the shared-memory in another sever, then with the application for the read operation from the replica of the shared-memory, the update is first performed on the replica. After that the lock has been received by the process, the read operation is carried out. Finally, for any operation, even the read, in order to be done by the process, it has to receive the lock to be privileged to read the valid values. The important point in this model is that, by receiving the lock for the read operation, the process does not need it to be released after its termination.

\subsection{Client-centric consistency model}\label{sec.ClientCentric}

\subsubsection{Eventual consistency}

This consistency model has a novel view to certain categories of distributed systems, where the update process thanks to its unconcurrency is rather easy. A great number of inconsistencies could be neglected thanks to this model in a relatively inexpensive way. The level of consistency between the processes and their confidence is variant in this model. Most of the processes barely preform the update on the replica, that is why a small number of the processes preform the update. On this basis, the only case which must be analyzed more frequently are the read/write conflicts where a process which wants to update a data-item while another process wants to read the same data-item simultaneously \cite{Tanenbaum2007}. In this case the update by this model is preformed by a lazy fashion like the lazy release consistency model \cite{Keleher1992a}. In this regard, the system tolerates a high level of inconsistency.  If there is no new updates after a long period of time, the system or the replicas will gradually become consistent \cite{Saito2005, Cachin2009}. Storage systems like Casandra \cite{lakshman2010cassandra}, Google Big Table \cite{chang2008bigtable}, and Amazon Dynamo are guaranteed in large scale service providing. For example, Amazon Dynamo chooses the eventual consistency \cite{brutschy2018static} in which each data-item in the replicas is synchronized gradually and decreases the synchronization overhead. The eventual consistency model \cite{Vogels2008} in case of the absence of the new updates for specific data, then they will receive updates from the nearest place where the data are reachable the latest update is read. However the read operation across multiple objects may return a combination of old and new values (non-integrated robustness values) \cite{Lloyd2011}. Most of the cloud storage systems rely on some of the eventual adaptation changes \cite{Kossmann2010}. The eventual consistency model thanks to the catch-all phase \cite{Li2012a}, shows the convergence for some of the replicas \cite{Lloyd2011}. By using a lazy fashion, it ensures the convergence of all replicas in the same way over time. This model by using the asynchronous lazy replication shows a better performance and faster access time to the data \cite{bouge2013managing}. However, it does not guarantee the same order of updates \cite{Bernstein2013a}. Epidemic replication is often applied to execute this consistency model \cite{Cooper2008}. Write operations which are preformed through the server are recorded in a file on the storage system, the general order of operations are acknowledged by the stamp, and the identification/ID and servers are then publicized \cite{Petersen1997}. In a system with eventual consistency, the server can utilize any kind of admit and verify method for write operations \cite{Terry1994c}. Finally, the two monotonic read and read your write consistencies, are the two desirable client-centric models in the eventual consistency, but not always required for this model \cite{Vogels2008}.

\subsubsection{Monotonic read consistency}

This consistency model is an example of the client-centric consistency models. When a replica requires data consistency, the latest changes on the data-item are sent to the demanding copy by the replica. The consistency model ensures that if a process observes a value in a certain time, then it would not see its previous value. An example in this respect would be the distributed email data bank where each user, has a mailbox in multiple replicated machines. As an example, in a distributed email data bank, each user has a mailbox in multiple replicas. Emails could be added anywhere to the mailbox. Only when a replica requires data consistency. Those data are sent to the demanding replica \cite{Tanenbaum2007}. Each session, determines the consistency insurance domain which could be the monotonic read consistency. One of the criteria to insure this consistency requires that the client keeps the previous session state \cite{Terry1234c, Bernstein2013a, Bermbach2013, Zhu2010, Liu2014, Bermbach2014, bouge2013managing}. These sequential read operations reflect a large set of write operations \cite{Terry1994c}. If the sequential read operation are preformed by a client, the updated content will be returned significantly \cite{Saito2005}. We have expressed the operating behavior of this consistency model in \ref{rule.09} \cite{brzezinski2004session}.

\begin{matriz} \label{rule.09}
	\begin{split}
		\forall_{C_i} \forall_{S_j}\begin{bmatrix}
			r(x)a \pdc r(y_i)b \Rightarrow w(x) a \xrightarrow{S_j} r(y)b
		\end{bmatrix}
	\end{split}
\end{matriz}

\begin{figure}
	\includegraphics[width=\columnwidth]{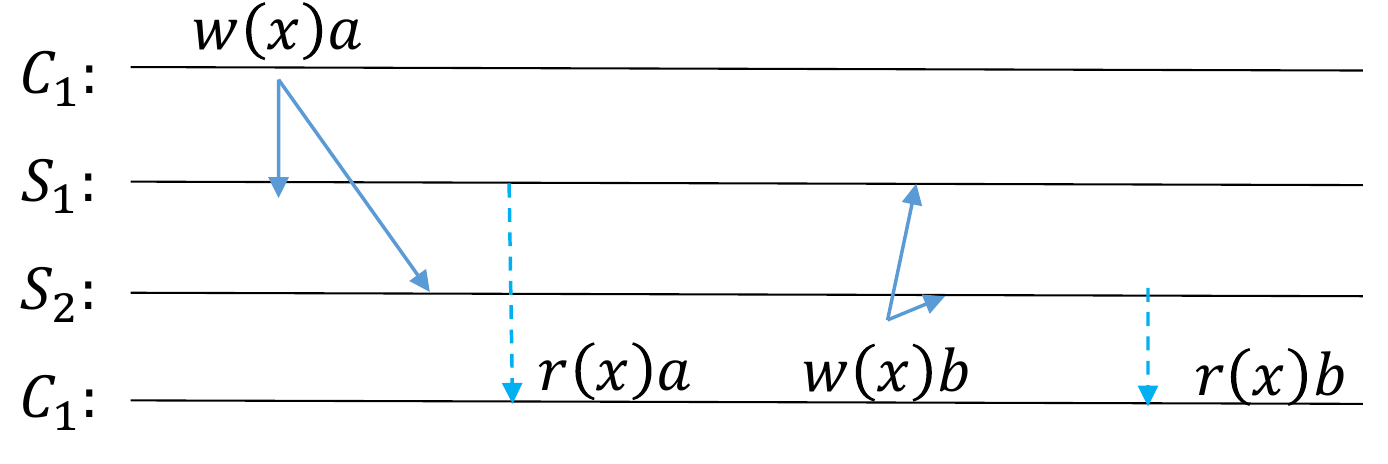}\\
	\centering{
		\caption{The behavior of the processes on the data-items based on the monotonic read consistency.}
		\label{fig.BehaviorMR}
	}
\end{figure}

In the monotonic read consistency model, if the client has read a value that from the memory, then it would not see any other values which have been written prior than that. In other words, by preforming the read operation by the client $ C_i $ a valid value is read from the replica on which all prior changes have been affected. As a result, with respect to \ref{rule.09} the read operation $ b $ is valid when the client $ C_i $ by having access to the replica existing in server $ S_i $ has executed all the changes on the object replica. Otherwise, it will face violations in the shared memory using the monotonic read consistency. To have a better understanding of this model, its behavior is depicted in Fig.\ref{fig.BehaviorMR} 

The behavior of the monotonic read consistency model on the shared memory is introduced as the client $ C_i $, can read the valid value b, when by having access to each of the replica in the servers $ S_1 $ and $ S_2 $ the whole prior operations have been executed on the replica.

\subsubsection{Monotonic write consistency}

The monotonic read consistency model is another variant of the client-centric consistency model. In this model is propagated with a correct order in the whole replicas of the memory. The write operation by the process on the data-item x is acceptable when the previous write operation is executed by the same process on the replica. Like the library of the software which needs to replace one or multiple functions in order to be updated. The important point is that in the monotonic write consistency, the write operations are done with the same sequence that have been started \cite{Tanenbaum2007}. The write operation by a client on a replica is ensured when the whole previous write operations are recorded by the same client on the same replica \cite{Saito2005}. This consistency model propagates the write operation with respect to the priorities between the operations \cite{Terry1994c}. Ultimately, it requires that each write operation be visible in order of presentation. Any sequence in the transactions like the Read Uncommitted is based on the priority specified by the global observer \cite{Bernstein2013a, Bermbach2013, Zhu2010, Bermbach2014, bouge2013managing, Bailis2013}.

\begin{matriz} \label{rule.10}
	\begin{split}
		\underset{C_i}{\exists}\begin{bmatrix}
			w(x)a \pdc w(y_i)b \\
			\Rightarrow \forall_{S_j} w(x) a \xrightarrow{S_j} w(y)b
		\end{bmatrix}
	\end{split}
\end{matriz}

The behavior of the monotonic write consistency according to \ref{rule.10} is expressed as follows: the write operation of $ b $ executes correctly when the write operation of $ a $ is preformed prior than that. In other words, when the client $ C_i $ writes the valid value $ b $ on the replica $ S_i $ that the operation of $ a $ on the same replica in the server $ S_i $ is done prior than that

\begin{figure}
	\includegraphics[width=\columnwidth]{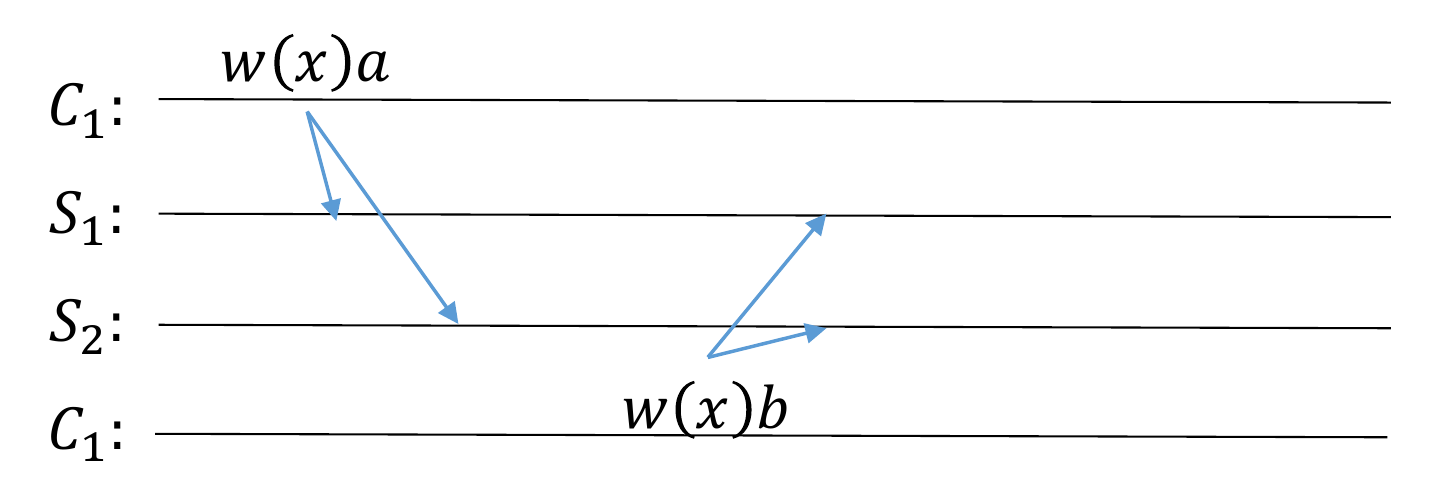}\\
	\centering{
		\caption{The behavior of the processes on the data-items based on the monotonic write consistency.}
		\label{fig.BehaviorMW}
	}
\end{figure}

This consistency model which is introduced in Fig.\ref{fig.BehaviorMW} expresses that the client $ C_i $ preforms the write operation of the valid value $ b $ on the data-item x if all previous write operations on that item are executed on the server $ S_i $.

\subsubsection{Read your write consistency}

In this model, the write operation is performed by the client on the data-item x which is always visible in succeeding read operations for the same client on the data-item $ x $. This means that write operations are completed before the next read operation is preformed by the same client. This consistency model could be specified as follows:

\begin{itemize}
	\item The data-access time is prolongated (such as password change).
	\item Similar to read-only consistency, but with the difference that the consistency of the last reading operation is determined by the latest client writing operation.
\end{itemize}

In this model, the new written value is read by the client instead of the previously written one \cite{Bernstein2013a, Bermbach2013, Zhu2010, Liu2014},  \cite{Bailis2013}. This model requires that each write operation be visible in order of execution. Any sequence in transactions such as the read uncommitted is based on the priority specified by the global observer \cite{Bernstein2013a, Bermbach2013, Zhu2010, Bermbach2014, bouge2013managing, Bailis2013}, and the read operation reflects its previous write operations \cite{Terry1994c}. Web page update could be named as one of the applications of this model. If the editor and browser are in the same program, the contents of the cache will be invalid when the page is updated, and the updated file will be downloaded and displayed \cite{Tanenbaum2007}. Read your write consistency can ensure editor and browser performance. We have described the behavior and mode of operation of this model according to \ref{rule.11} \cite{brzezinski2004session}.

\begin{matriz} \label{rule.11}
	\begin{split}
		\underset{C_i}{\forall} \underset{S_j}{\forall}\begin{bmatrix}
			w(x)a \pdc w(y_i)b \\
			\Rightarrow w(x) a \xrightarrow{S_j} r(y)b
		\end{bmatrix}
	\end{split}
\end{matriz}

\begin{figure}
	\includegraphics[width=\columnwidth]{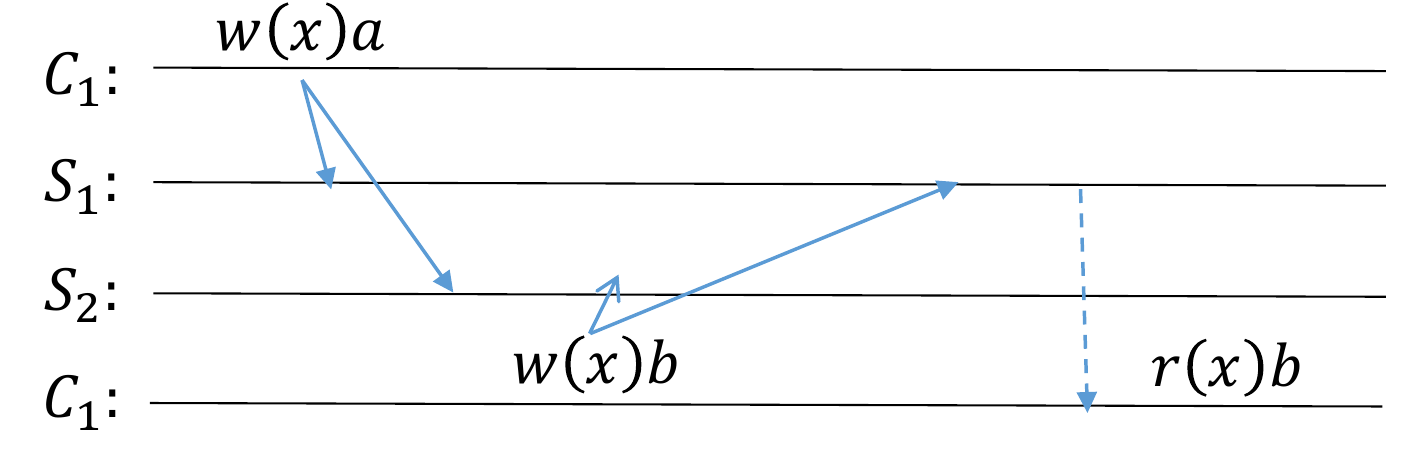}\\
	\centering{
		\caption{The behavior of the processes on the data-items based on the read your write consistency.}
		\label{fig.BehaviorRYW}
	}
\end{figure}

The read your write consistency model is expressed according to \ref{rule.11} as follows: the client $ C_i $ can read a valid value b when the write operation a is previously executed on the replica existing in the server $ S_j $. Where b is the valid written value by the client $ C_i $. In order to have a better understanding of this model, Fig.\ref{fig.BehaviorRYW} illustrates the way this model works.

The behavior of this model in accordance with Fig.\ref{fig.BehaviorRYW} is such that the client can read the new value b as a valid result, before the execution of the write operation b by the client $ C_1 $ on the data item in the server $ S_i $ , all write operations on that particular data item are executed in the existing version of the server. Otherwise, it will encounter inconsistency in the shared data repository.

\subsubsection{Write follow read consistency}

In the write follow read consistency which is also known as the causal session \cite{brzezinski2004session}, the update propagations are based on the latest read operation. The client can write on the data-item $ x $, when the latest value of data-item $ x $ is read by the same client. For instance, in the twitter a client can post a re-tweet on a post that he/she has already seen. With the read operation, this model verifies the previous dictating write operation of the client and ensures the new write operation \cite{Saito2005}. This model is related to the happen-before relation \cite{Lamport1978a} which is proposed by Lamport \cite{Bermbach2013, Zhu2010, Bermbach2014, bouge2013managing, Bailis2013}. \ref{rule.12}, show the behavior of this model:

\begin{matriz} \label{rule.12}
	\begin{split}
		\underset{C_i}{\exists} \begin{bmatrix}
			r(x)a \pdc w(y)b \\
			\Rightarrow \forall S_j w(x) a \xrightarrow{S_j} w(y)b
		\end{bmatrix}
	\end{split}
\end{matriz}

\begin{figure}
	\includegraphics[width=\columnwidth]{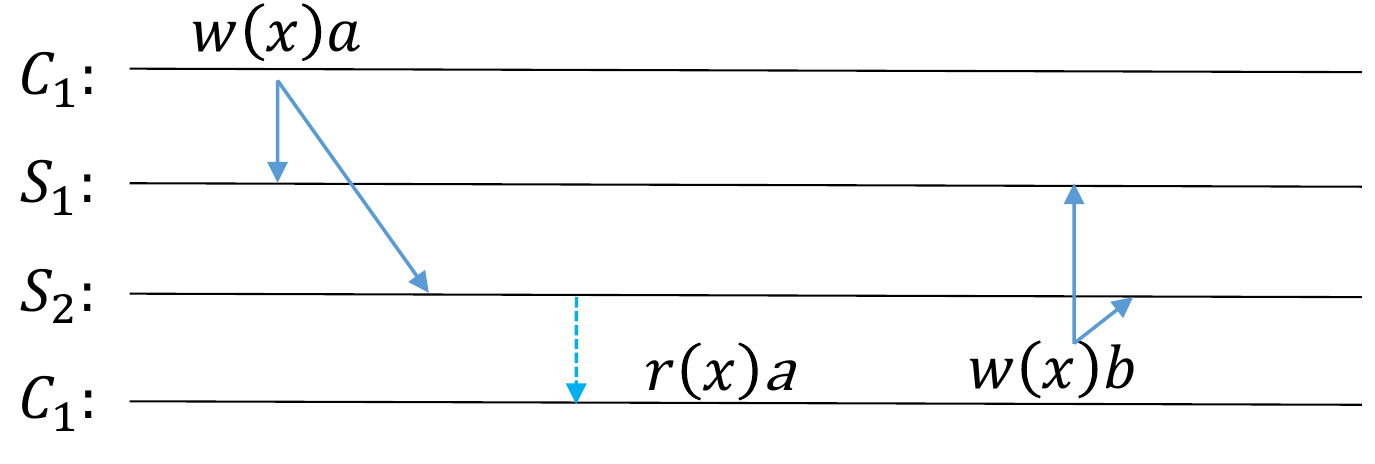}\\
	\centering{
		\caption{The behavior of the processes on the data-items based on the write follow read consistency.}
		\label{fig.BehaviorWFR}
	}
\end{figure}

As it can be seen in Fig.\ref{fig.BehaviorWFR}, if the client $ C_i $ read the value $ a $ from the data-item $ x $ on the server $ S_j $, then this value is written on the same server and then writes the new value $ b $ on data-item $ x $ on server $ S_j $ and as a result the write follow read operates correctly.

\section{Novel consistency models}\label{sec.Novel}

Over the course of the years, the essentialities of distributed systems, especially cloud environments, have changed. In spite of the malicious cloud system, the necessities such as security and reliability, or even the need for greater convergence in operations, have become apparent in these systems. With this in mind, researchers have introduced new consistency models to address these needs and challenges, which are presented later in this section. 

\subsection{Fork consistency}

This is a client-centric consistency model, which expresses the strongest concept of data integrity without the presence of the online servers and reliable clients. If the service prevent a client from viewing another  client's update, then the two clients will never see the update information of each other. This model was originally introduced for file systems which conceal client' operations from each other. According to this model, if we divide the clients into two groups each of these groups cannot see the operations of the other and vice versa \cite{Brewer2012}. Making decision on whether to use the accessibility or the peer-to-peer communication as the partitioning criterion has a great impact on the consistency of the model \cite{Li2004}.

Fork consistency increases the concurrency of the operation on the shared data in the distributed systems and is introduced to protect the client information from the malicious systems. This model is presented for unreliable storage systems and their clients who are not in direct contact with each other. In fork consistency, when an operation is observed by the clients, they have the same view of the previously executed operations. When a client reads a value written by another client, then write consistency of the writer is guaranteed for the reader.
SUNDR is a network file system protocol with security used in unreliable data storages, the purpose of this protocol is to ensure the fork consistency \cite{Cachin2011, Mazieres2002}. This consistency model is guaranteed, when the following three properties are considered and implemented by the data storage \cite{Li}.

\begin{itemize}
	
	\item Function Verification: This property includes a list of correct operations sent by a healthy client to the server, which is called the issue time.
	
	\item Self-consistency property: The characteristic of this property is that every healthy client sees all his previous operations. This property assures that it is consistent with its operations. For example in a file system, the client always sees its own write operations.
	
	\item Unconnected Feature: Each list is the result of the correct operation of a healthy client, with other clients having the same view of the operation of that healthy client.
	
\end{itemize}

This consistency model has been introduced with three properties, if the third property of this model is modified to enhance the precision and improvement of the system inspection to the join-at-most-once property, then this model will be converted to the fork* consistency (FC*). As stated before, the advantage of this model is to increase the precision of the system inspections. This model is also used for unreliable client storage systems \cite{Li}. Among the applications used in this model, the SPORC is presented as a framework that complements the benefits of fork* consistency and operation transitions . This model operates unlocked simultaneously and automatically eliminates the conflicts resulting from the operation's consistency \cite{Feldman2010}.

Another type of fork consistency model is the fork sequential consistency (FSC)\cite{Oprea2006}, which is a hybrid consistency model. When a transaction is viewed directly or indirectly by multiple clients, all clients see the entire previous operations according to their event history \cite{Cachin2009}.

For example, when a client reads the value written by another client, the reader will make sure of the consistency of the value written by the writer. In this model, the clients have complete satisfaction of the order in which the operations are performed. However, in the FSC, there are no guarantees that the operation of all clients are atomic. In this case, in addition to making sure of the sequence of the operations done by the healthy client on the healthy server, the atomicity the operation must also be guaranteed.For this reason, another hybrid consistency, called the Fork Linearize Consistency (FLC) \cite{Mazieres2002}, has been introduced.

When the accuracy of the server is determined, the atomicity of all client operations on the server is guaranteed by this model \cite{Cachin2011}.
This means that all clients view the other clients' operations in a consistent and uniform manner.
If the server is accurate and reliable, the service must ensure linear consistency, and if it is not trustful, it ensures fork consistency, and only when the time stamp is used this protocol is applied correctly to preform the concurrent operations. If the server is correct, this protocol will execute all operations in serial and lock-step procedures and will not permit the execution of the operation.

Verification integrity and consistency object’s system (VICOS) is a lever for ensuring linear fork consistency for the application-oriented object-oriented storage system. In this case, the storage system should be transparent \cite{Brandenburger2015a}. The cloud storage system can be well-adapted to the linear-line consistency model, but it does not tolerate any malicious or problematic behavior from the client or the server.

Depot is a storage system with minimum reliability for which the Fork-Join-Causal consistency model has been presented \cite{mahajan2011depot}. The system ensures two previously introduced properties with the presence of the malicious nodes and keeps the updates with respect to stability, readability and restoration consistent. The fork-join-causal consistency is also called the View-Fork-Join-Causal \cite{Mahajan2011b}. This model minimizes the number of acceptable forks to run. This model has the strongest accessibility and data semantic convergence at the presence of the malicious nodes.

\subsection{View consistency}

In system with distributed proof structure, propagation of the information policy is preformed to have thorough knowledge on the complete structure of each proof tree. This knowledge for each of each proof tree is completely unexpected. This model, ensures the consistency in which the data are encrypted by the certificate authorities.

Using this model, the constraints of consistency could be executed in the proof tree systems. In this model, it's unlikely to show the details of the proof completely for issuing privileges and evidences. It is likely that this type of distributed proof system is similar to that used in computing and the network environment of sensors. This model is one of a variety of data-driven consistency models \cite{Lee2007, Lee2010}. But before looking at how this model operates, a definite definition of observation must be provided. View means a view of each set of real-world identities that are not more than one pair for each id pair $  <id, e> $. The observance of the entity $ e $ is defined the system as a set of local observations. With respect to the fact that each observation only includes the local ones, a detailed snapshot of system status is unlikely. Therefore, the consistency level of these observations is of great importance. The system with the view consistency would be consistent if and only if the system has a valid view of the stored data status at specified intervals. However, three levels of the view consistency related to proven distribution protocols are described to be applied in the system \cite{Lee2007}.

\begin{itemize}
	\item  Increment consistency: the most introductory definition of the view consistency is the increment consistency which is more frequently used than the other consistency levels. The increment view consistency is that any operation during the construction of a related proof tree is valid in some points. A view of the completion time interval increment consistency of proof tree and the receipt of the request submitted by the applicant. In this case the increment consistency runs correctly. In theory 1, the protocol of the distributed proof Minami-Kotz, always uses the increment consistency when the authorization policies are being evaluated. In fact, the existing distributed construction protocols, use the increment view consistency when making a decision about the privileges. This phenomenon leads to various safety violations. Therefore, the increment consistency  is not ensured due to the overlap of validation intervals stored in the system.
	
	\item Query Consistency: the next level in view consistency is the query consistency where the whole operations used to create the distributed proof are valid when the triggering query the proof creation are done simultaneously. If the privilege policy is satisfied of using the query view consistency, then this policy is satisfied in the creation environment of the distributed proof consistency using a centralized proof framework to support the transaction evaluations.     
	
	\item Interval Consistency: another level of the view consistency is the interval consistency, in which the operation of the model is completely accurate if and only if the state of each multiple operation which consists of the validity of two accurate time intervals is encrypted. Nevertheless, some researchers have introduced a fourth state in order to execute this model more accurately \cite{Lee2010}.

	\item Sliding Window Consistency: although the concept of the interval consistency in some cases has been defined as the strong consistency. However, there might be situations where the observation face consistencies for multiple times and lead to multiple interrupts in providing services. For instance, in the pervasive computation environments with frequent text alternations, the accepted validity state of the operation might alternate. A share of the weakness in the triggering query consistency is related to the noncontinuous validity check of the view consistency. The entities might be interested to execute a level of the view consistency (somewhere between the query consistency and the time interval). A level between these two consistencies is called the sliding window consistency. This level of consistency requires a sequence of observations recorded about the validity of the applied operations to make the distributed proof. As a result, the algorithm used to execute this mode of consistency cannot utilize the ``build and credit'' strategy used to execute the interval and query consistencies. Yet, the algorithm in order to limit the sliding window consistency has to execute the instance of the proof tree for a number of times in order to evaluate all of the consistency conditions.
	
\end{itemize}

\subsection{Multi-dimensional consistency}

Multi-dimensional consistency \cite{Yu2002}, is a data-centric consistency which has introduced a unit called ``conit'' as a unit for consistency with a three-dimensional vector. This model handles the deviations like the sequential or numeric errors from the linear consistency model. Yet, the numeric error is often non-executable and in terms of definition has overlaps with old and sequential errors \cite{Bermbach2013}. However, there are no errors to be disregarded in this consistency \cite{Tanenbaum2007}. Multi-dimensional consistency, is a model for the distributed services. TACT, is a middle-ware layer which executes the continues consistency based on the conit unit using the three aforementioned features among the replicas \cite{Yu2002}.

\subsection{VFC3 consistency}

Nowadays, the dependency to the data stored in the cloud data centers has surged dramatically all over the world. Different replication protocols are applied in order to achieve high accessibility and performance and to guarantee the consistency between the replicas. By using the traditional  models, the performance of the consistency might decay. Therefore, most of the large scale data centers take the declination in consistency brought about by the decrement in the time delay for the end-users into consideration. Generally, the level of consistency is reduced by those cloud based system which give privileges to the stale data at random to stay in memory for constant periods of time. Moreover, the behavior of such systems causes ignorance to the data semantics. This behavior, the necessity to combine the strong and weak levels of  view consistency is felt completely. In order to have better consistency in dealing with accessibility the VFC3 model of consistency is proposed \cite{Esteves2012}. The novel consistency model is used to replicate data in data centers under the library framework in order to increase the consistency level and is based on the three-dimensional vector of time, sequence and value related to the data objects. Each of the dimensions is a an scalar which shows the maximum level of discretization from the limitations of the consistency. By considering the following dimensions, this model provides the consistency:

\begin{itemize}
	\item Time dimension: denotes the maximum time that a replica having the latest value cannot be updated.
	
	\item Sequence dimension: shows the highest update frequency which can be granted for an object without considering the replicas.
	
	\item Value dimension: represents the maximum proportional difference between the data content of the replica to a constant value.
	
\end{itemize}

\subsection{Timed consistency}

Sequence and time are two different dimensions in the consistency of the shared objects in the distributed system and specially cloud based systems. One of the ways to avoid the in consistency between the operations is an effective and fast method in order to take the effects of an operation in the system. The sequential and causal consistency models, do not take the specified and valid time of the executed operation into consideration. The timed consistency \cite{Cooper2008} is a data-centric consistency model. In this model if an operation is not executed in time $ t $, the other nodes must be observable in time $ \Delta + t $. In some cases, this model is also referred to as the delta consistency. This model describes a combination of the sequential and old inconsistencies. In other words, the time interval in the system is terminated and the system reaches a steady state. This means that the minimum guarantee of a specific consistency model is granted in a fixed time period $ \Delta t $. If the replica during the time period fails to synchronize, then the desired item would not be accessible until the time that the replica is consistent again. This model is frequently used in order to guarantee the service level protocols and also increase the clearance of the operations between the consistency and accessibility \cite{Bermbach2013}. Some other mentioned models such as the sequential and causal consistencies are the two models that have presented the hybrid consistency. In the timed sequential consistency model, the time criterion is involved in determination of the priorities in the execution sequence of the operations. If the operation $ e_1 $ in time $ t $ and operation $ e_2 $ are executed in time $ t + \delta $, then the whole nodes or processes observe the operations according to the execution time sequence \cite{Torres-Rojas1999}. As mentioned before, the timed causal consistency model is also another hybrid consistency model which is involved during the execution process of the timed consistency. In this model, using the operation event time, the causal consistency is executed with more validity. Whenever the operation $ e_1 $ is executed in time $ t $ and be the cause to the operation $ e_2 $ in time $ t + \delta $, the causal relation $ e_1 \rightharpoondown e_2 $ is valid \cite{Torres-Rojas2005}.

\subsection{Coherence model}

This model is one of the data-centric consistency models which provides and guarantees the sequence of operations on all data items \cite{Bermbach2013, Tanenbaum2007}. For instance, in the causal consistency, two updates on two different data items by a client has to be executed according to the correct serial sequence. Besides, a storage  system can reach the steady states when all the replicas are the same from all data items. Reaching a steady state is not feasible according to the size of the data store. Coordination between the updates in short period of time on a great number of servers is an extremely hard task to do. Thanks to the scalability feature, mostly the consistency model is insured to be executed on each data store consistency key. The introduced models are called the eventual coherence, causal coherence and sequential coherence. 

\subsection{Adaptable Consistency}

Adaptable or rationing consistency \cite{Kraska2009} is executed as soon as the data-items are clustered proportional to their importance e.g., in on-line based stores the credit cards use this model of consistency on different types of data-items i.e., $ A $, $ B $, $ C $.

Although the $ A $ and $ C $ data-items receptively apply the linear and eventual consistency models, data-item $ B $ calculate the continues alternations of this consistency based on the inconsistency cost function. Whenever, the cost of inconsistency exceeds the cost of inaccessibility or high latency, then linear consistency is performed on] data-item $ B $ \cite{Bermbach2013}. For instance, cloud-based rationing consistency is executed through GARF library \cite{Abstract}.

This consistency model with self-adaptability \cite{Chihoub2012} in the cloud environment in selected based on the consistency cost \cite{Chihoub2013}, it privileges the client to specify the maximum stale read rate or the consistency cost according to the Service Level Agreement (SLA) \cite{Bermbach2013}. This  model based on the type of cost and data-item present different levels of consistency. The RedBlue consistency like adaptable consistency provide two discriminant levels of consistency based the type of operations \cite{Bermbach2013, Li2012a, Sivasubramanian2012b}.

\subsection{RedBlue Consistency}
	
The ReadBlue consistency \cite{Li2012a, Sivasubramanian2012b} is presented to increase the replication speed in the distributed systems. In other words, the increment in speed suggests that whenever the client send a request to the server, it receives its response in a short period of time. Eventual consistency \cite{Vogels2008}, by reducing the synchronization among the nodes or sites processes the local operations with a faster speed. In contrast, the linear and sequential consistencies as they have high rates of communication in their synchronization process among the nodes, they prohibit the agility in processing of the operations.

The RedBlue consistency categorizes the operations based on the type of execution into two sets of red and blue operations. The execution order of the blue operations can be different from one site to another, whereas the red operations should be executed for all sites equally the same. This consistency model consists of two parts: 1. the order of RedBlue which specifies the order of the operations, and 2. a series of local serializable operations which have causality relation between each other. The causality relation in this model records these realtions in other sites by assuring that the dependency of the operations are recorded in the main site and guarantees them.

In definition, the sequence of operations in specific sites are processed locally. In such system each operation with a red label is executed in a serializable order \cite{Hadzilacos1987} while each blue labeled operation is performed like the eventual consistency \cite{mahajan2011depot, Lloyd2011, Terry1995a}. These labels specify the category of operation. The execution of this model makes sure of the absence of the inconsistency in the attributes of the application.

Finally, the whole replicas are converged. The operation in this model has no effect on the replacement of the blue operations. On this basis, a method is introduced in order to increase the feasible space to execute the blue operations by dividing them into two phases of the operation generator and shade. The operation generator merely finds the alternations in the main operation. In the recognition phase, the blue and red operations are identified. However, in the shade operations the identified alternations are executed and are replicated on all sites and the operations are the shade, blue or red exclusively. 
	
\section{Challenges and issues}\label{sec.EvaluationParams}

\begin{figure*}
	\includegraphics[width=\textwidth, scale = 1]{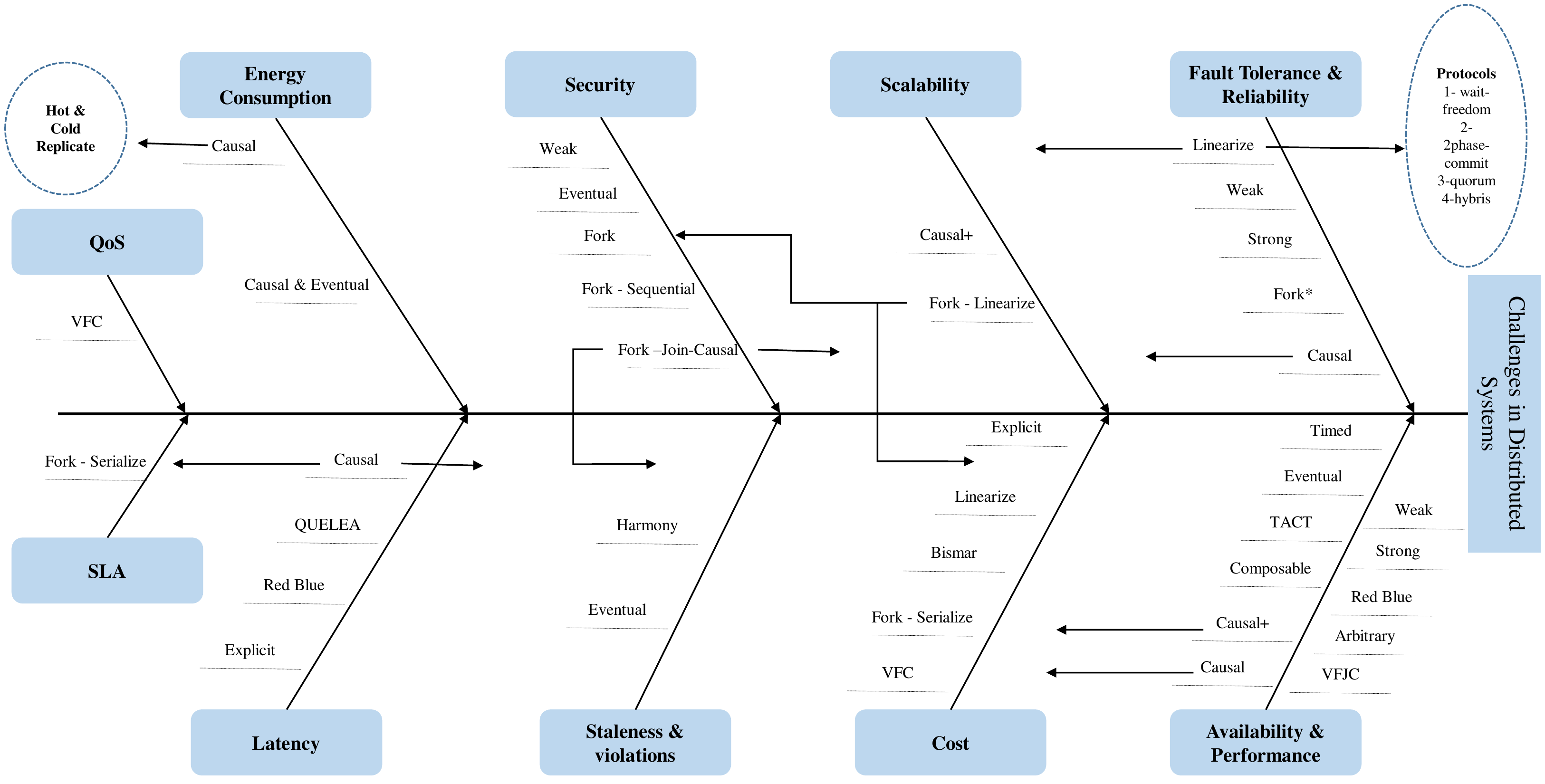}\\
	\centering{
		\caption{Classification of different types of consistency based on the introduced challenges in the distributed systems.}
		\label{fig.challanges}
	}
\end{figure*}

Some of the proposed consistency models such as linearize, fork-linearize, fork-join-causal, causal, and causal+ consistency which are proposed by the researchers to cope with different challenges. As you can see in Fig. \ref{fig.challanges} a number of consistency challenges in the distributed systems are projected. Specially, the linearize which is utilized to increase the fault tolerance and reliability by applying the protocols such as: wait-freedom, 2phase-commit, quorum based replication protocol and hybirs. The trade-off between consistency and availability is the most important challenge on which researchers have conducted a great number of researches during recent years. After that, the reliability, cost, and security in the distributed systems have been focused by the scientists. In our research we have proposed a new classification which discusses the challenges covered by consistency models. Data consistency results in an increment in security and user reliability in the system. By emplacing the replica at the nearest point to the client latency and response time are expected to be reduced. Therefore, it brings about more accessibility to the data. Data-centric consistencies by executing the partial replication protocols not only increase the scalability and availability, but also reduce the consumed energy. The trade-off between consistency, violation, and staleness is that by increasing the consistency the others deteriorate. In case the distributed system is based on the cloud, the Quality of Service (QoS) and Service Level Agreement (SLA) are the challenges. The decrement in violation and staleness results in an increment in the QoS in the cloud. 

\subsection{Reliability and fault tolerance}
In order to increase the reliability, the data are replicated in the distributed system. Fault tolerance can be in variant forms such as using the other replicas in case of the failure of the local replica, keeping several replicas to maintain against the distorted data \cite{Tanenbaum2007}, tolerance of read aborts and data deletion by the malicious client \cite{Goodson2002}, and failure in message sending from the server to the client or vice versa and also message loss, etc \cite{Lamport1982}. GARF is an object oriented system which has proposed three class of Slow, Parm, Causal, and weak consistency in order to increase the reliability in the system \cite{Abstract}. Message passing systems consent on two third of the values of replicas in order to increase the fault tolerance \cite{Lamport1982}. However, researchers have proposed systems with stronger consistencies e.g., linear.

Linear consistency using the 2phase-commit, wait-freedom, quorum-based, and Hybris protocols performs as follows:

\begin{itemize}
	\item In 2phase-commit avoids access to erasure code and delete them by malicious client \cite{Goodson2002}.
	
	\item In wait-freedom prohibits the execution of the read or write operations of the erasure code in heterogeneous systems with malicious servers or clients \cite{Goodson2003, Goodson2004}.
	
	\item In quorum-based in order to increase the fault tolerance a consensus must achieved on the data contained by the replicas \cite{Kong2005}.
	
	\item In Hybris, it provides linear, although public cloud comes with the eventual consistency \cite{Dobre2014}. Hybris also tolerates the connection loss with the cloud besides it coherently replicates the metadata of the public cloud in private cloud.
\end{itemize}

However, the performance of these protocols in Byzantine Fault Tolerance (BFT) with more the $ f $ faulty replicas is not acceptable. As a result of this deficiency the BFT2F which is an extension over BFT has been proposed in order to append the linear to the BFT. This algorithm shows a more promising performance with more than $ 2f $ faulty replicas. This prohibits the malicious servers from responding to client requests and depicts the inconsistencies in the system \cite{Malugu2015, Kwon2014}. The Shared Cloud-backed File System (SCFS) provides the strong consistency along with the semantic POSIX on the cloud which uses the eventual consistency and is able to discriminate between the malicious and benign Cloud Service Providers (CSPs) \cite{Bessani2014}. A framework using the Fork* in order to communicate the faulty servers and tolerate the low speed or lost network connectives is provided which solves the conflict results using the Fork*\cite{Feldman2010}. In contrast, Eiger, a geographical distributed storage system, for which the causal consistency is provided in order to tolerate the network partition disconnectivities between the data-centers \cite{Lloyd2013}. In recent years, thanks to the disk which come with the RAID technology, the state machine replication (DARE) with high performance prohibits the total network connection loss \cite{Poke2015a}. 

\subsection{Performance and availability}
Placing a copy of data near to the process or the client which uses it leads to an improvement in performance in distributed systems \cite{Tanenbaum2007, Lady2014}. GARF by proposing the three classes of Atomic, Sequential, and CAtomic provides the strong consistency criteria which satisfies the availability \cite{Abstract}.

The trade-off between consistency and availability indicates that the system in some time intervals by tolerating the inconsistency causes an increment in the data availability. With respect to this trade-off the weak consistency on the structured log file replicas results in an increment in availability \cite{Spreitzer1997}. Also, the timed consistency is a weak consistency which defines a threshold for the access time. Access time, is the reasonable time for data access \cite{Torres-Rojas1999}.

The functionality of a system is differs based the types of consistencies which are applied in it. TACT is a middle-ware in distributed systems which using numerical staleness ordering deviations are the criteria which selected arbitrarily in the continues consistency based on Conit in order to improve the system performance\cite{Yu2002}. In contrast, the composable consistency, five criteria of concurrency, consistency, availability, visibility and isolation are presented the combination of which results in the favored consistency and improves the functionality of the system \cite{susarla2003composable}. However, the arbitrary consistency using the partial replication, a mechanism which stores or replicates the data on each storage node and improves the performance of the system \cite{belaramani2006practi}.

High availability could be achieved by sacrificing the consistency. ZENO, a state machine replication, in BFT in order to achieve high availability replaces the eventual consistency with the linear. This machine, as long as the network connectivity is approved, with a consensus over the latest update between the replicas serves the client requests \cite{Singh2009}.

The View-Fork-Join-Causal (VFJC) consistency with respect to the Consistency Availability Convergence (CAC) theorem in order to converge and access the replicas is presented \cite{Mahajan2011b}. Eiger using the causal consistency grants the  ease of access for the clients \cite{Lloyd2013}. However, in comparison with the causal and VFJC, the causal+ consistency in chain reaction storage systems with network partition is satisfied with high availability. This consistency increases the convergence among replicas \cite{Almeida2013}. 

The causal consistency using the Consistency Availability Partition tolerance (CAP) theorem stands the network partitioning and is satisfied with the availability \cite{Brewer2012, Birman2012, Shim2012}. However, the PACELC \cite{abadi2012consistency} which is forked from the CAP theorem, in case of network partitioning, illustrates the trade-off between consistency and availability \cite{Abadi2012, Diack2013}.

In recent years, a new type of consistency known as the RedBlue is proposed which divides the operations into two subcategories of local operations (Blue) and global operations (Red). Eventual consistency with lazy fashion replication enforces the Blue operations so that the clients have more accessibility to the data in their local operations \cite{Li2012a}.

The daily increment in client demands for the high accessibility to the data has made the researchers to propose the eventual consistency with high availability \cite{Brewer2012, Birman2012, Shim2012, Vogels2008}. However, in recent years, for the JSON data model and storage systems with high availability, the eventual consistency with convergence and liveness property is proposed \cite{kleppmann2017conflict, Attiya2017}.

\subsection{Scalability}

Scalability is one of the most important points of interest for the researchers in the distributed systems. Scalability can viewed form different points such as increment in number of the servers, data centers, clients as wells as their requests, and the expansion in the geographical area \cite{Tanenbaum2007}. Expansion in size occurs as soon as a great number of processes need to have access to the data stored in a server. In recent years, researchers have analyzed the trade-off scalability and consistency in the geographical distributed systems which prohibits the inconsistency in case the expansion in the scalability. Cassandra is an scable distributed storage system which provides multiple consistencies \cite{lakshman2010cassandra}. However, the other storage systems such as COPS and Chain Reaction are scalable in terms of an increment in the number of the servers in each cluster and only provide the causal+ consistency \cite{Almeida2013, Lloyd2011}. Moreover, besides being scalable, COPS provides data access as well as convergence among replicas. In terms of the cloud computing scalability could be emerged as an expansion in the number of clients, servers, and CSPs which prevents the bottleneck phenomenon the systems \cite{Lloyd2011, Kraska2009, Cachin2011, popa2011enabling, Kim2015a}. Yet, recently a service named SATURN is proposed which provides partial replication and causal consistency. This highly scalable ignorant to the number of clients, servers, data portioning, and locality \cite{bravo2017saturn}.

\subsection{Cost/Monetary Cost}

There exists a trade-off between consistency and cost. Data is replicated in the distributed systems in order to increase the performance of the system. However, update propagation is costly \cite{Tanenbaum2007}. The cost parameter itself includes the bandwidth for data transmission among replicas, data overhead, storage, monetary cost \cite{Vogels2008, Kraska2009} based on the cloud service provision. BISMAR has considered the cost efficiency in different consistency levels in cassandra \cite{Chihoub2013}. In recent years, some researchers have presented the causal \cite{Shen2015} as wells as liner \cite{Bessani2014} consistencies in order to reduce the storage space in the distributed systems. The fork-linearize \cite{Shraer2010}, fork \& serialize in cloudproof \cite{popa2011enabling}, $ VFC^3 $ \cite{Esteves2012}, causal \cite{Lloyd2013},\cite{Kim2015a}, causal+ \cite{Almeida2013}, linear with the hybris protocol \cite{Dobre2014} are the consistencies in order to reduce the bandwidth and network overhead. Explicit is an extended eventual consistency which is presented to reduce the synchronization cost of conflict operations and optimal storage in the geo-distributed systems \cite{Balegasa}. 

\subsection{Security}

With the emergence of different distributed systems such as grid computing \cite{foster2003grid}, cloud computing \cite{buyya2009cloud}, edge computing, fog computing, mobile \& mobile cloud computing \cite{Yang2017, roman2018mobile}, the accessibility of the malicious servers and clients to data is vital challenge in the distributed systems. Weak consistency using the protocols and system design has increased the data access security. Weak consistency has been proposed for the (a) structured log files of the replicas, (b) fixed servers which store the client operations using their digital signature (c) or audit servers which inspect the malicious servers \cite{Spreitzer1997}. Along with the provision of structured logfiles, Bayou's anti-entropy protocols for data transmission between the replicas and eventual consistency through digital certificates and trusted delegates  were provided to ensure the security over the insecure wireless communications or the Internet \cite{Petersen1997}. With the passage of time, Fork consistency and a variety of consistencies derived from it have been introduced to enhance the level of data security in dealing with servers and malicious clients. SUNDR is a distributed file system that, by Fork consistency and digital signature, automatically detects any maladaptive behavior in the server sectors \cite{Li2004} and ensures that clients can recognize any disintegraty and consistency errors, until they observe file changes made by another client \cite{Mazieres2002}.

Other protocols such as key distribution and file storage protocols are developed to generalize the fork consistency on encrypted distributed file systems to protect user data on untrusted servers \cite{Oprea2006}. Fork consistency in terms of performance has proven to be good for dealing with malicious servers, but it has brought about system failure in dealing with fixed servers. Therefore, researchers are looking for a combination of the fork consistency and other types of consistencies. Fork-linearize and fork-sequential are hybrid consistencies out of which the linearize or sequential consistency runs on the fixed server to ensure the client data and in case of the malicious servers the fork consistency is applied.

But, the main disadvantage of these two hybrid consistency models is that when they are faced with the fixed servers, they do not allow server operations to be executed in absence of the Wait-Freedom protocol \cite{Cachin2009}. Besides, Venus is  introduced as a security service to interact with untrustworthy servers, which combines three consistency models in its service. It also recruits the Fork-linearize hybrid consistency to deal with malicious servers and uses the eventual consistency for high data availability \cite{Shraer2010}. Furthermore, one of the biggest advantages of fork-linearize consistency is when the clients are not interacting with each other and want to ensure the integrity, and integration of the shared data on unreliable servers, this consistency will help them ensure that the operations of all clients are guaranteed on these servers \cite{Cachin2011}. Moreover, the fork-join-causal consistency in the Depot distributed system, not only guarantees the safety and biological characteristics for the faulty nodes, but also provides better data stability, durability, scalability, and accessibility by tolerating the network partitioning \cite{mahajan2011depot}.

\subsection{Staleness and violations}
Applications in different ways determine that which inconsistencies can be tolerated. By defining three independent paradigms, a general method for determining the inconsistency is presented \cite{Yu2002}: the deviation in numerical values between replicas or the deviation according to the order of the update operation indicates the degree of severity of the violation and the deviation in reading between the replicas shows the stale read. Distributed Depot system, using Fork-Join-Causal consistency, in addition to guaranteeing the stability and durability of data, reduces the stale read by causal consistency by tolerating network partitioning \cite{mahajan2011depot}. Harmony is the consistency proposed for the Cassandra, which supports several consistencies \cite{Chihoub2012}. This consistency is related to the application type of self-adaptive consistency. Eventual consistency which is devised in harmony has shown better performance in reducing the stale read time. Even the eventual consistency in comparison with some data-centric consistencies has also shown a better performance in reducing the staleness in reading \cite{Rahman2012}. With indirect monitoring of consistency, it is possible to calculate the optimal execution cost of consistency and examine the stale read with respect to the consistency behavior \cite{Bermbach2014}.

Or even by local and global replica inspections that reduce the severity of violations and the rate of reading slowly through consistency of readability and adaptability to read self-writings and global audits by causal adaptation \cite{Liu2014, Padmapriya2015, Kulkarni2015}. The rate of readability and the severity of the violation are two of the challenges that interact with the cost, quality of service, and service level agreement in distributive environments such as cloud. Reducing the aging rate and the severity of the breach reduces the cost of compensation for noncompliance \cite{Kraska2009, Chihoub2013} or even the cause of reducing bandwidth consumption or reducing data overhead in the network \cite{Kim2015a, Golab2011}. A service level agreement or quality of service is a challenge that has been introduced in recent years by providing consistency as a service. As stated earlier, the readability and severity of violations of the inconsistency determination axes indicate improved service quality or service level agreement based on the level of compliance promised \cite{Liu2014, Padmapriya2015, Kulkarni2015, Anderson2010}.

\subsection{Quality of Service (QoS)}

$ VFC^3 $ based on the functionality of the cache memory has resulted in an improvement in QoS. Data replication and storage in the cache memory in the nearest node to the client reduces the response time of the client request and bandwidth \cite{Esteves2012}. Furthermore, there exists a trade-off between the QoS, Violations, and Staleness where a decrement in the later ones results in an increment in the QoS \cite{Liu2014, Kulkarni2015, Anderson2010}. 

\subsection{Service Level Agreement (SLA)}

With the emergence of the cloud computing and the introduction of the consistency as a cloud service by the CSPs, the researchers confronted with a new challenge called the SLA. Recently, this challenge based on the least response time has been proposed \cite{Terry2013a}. Besides, a melioration in QoS of the consistency in the cloud brings about the promised consistency level in the SLA \cite{Liu2014, Kulkarni2015, Anderson2010}.

\subsection{Latency}

In geographically distributed systems several big data-centers are condensed in different spots all over the world. These large-scale data-centers provide computers with sufficient resources to serve many users. However, this concentration of the resources leads to a gap between users and resources. This, in turn, causes network latency and jitter \cite{roman2018mobile}.

Researchers use replication to reduce the data access latency to a local replica of the data in the nearest location to the client, this latency includes the duration of write or read, the transmission of messages between the client and server, and eventually the data access time.

Although eventual consistency is one of the weakest consistencies, it has alleviated the data access, write, and read latencies. Additionally, the Indigo middleware layer has provided an extended eventual consistency named Explicit on the storage system \cite{Balegas2015}. Similar this middleware layer, QUELEA is presented for the distributed systems with eventual consistency. However, QUELEA is a programming model that is able to detect the consistency features at the granular level of the application, which results in a reduction in latency \cite{Sivraramakrishnan2015}.

In distributed systems, full replication is applied to reduce the response time of the client requests. Yet, by introducing of the causal consistency and partial replication, not only the storage overhead and replica transference has decreased, but also the researchers have reduced the response latency to the client requests \cite{Shen2015}.

In recent years, the novel consistency has introduced the RedBlue, which besides to its other benefits plays the role of the eventual consistency to the local operations of the clients who are in Blue's operations to reduce access and response to client demand \cite{Balegasa}. However, by disseminating data by intelligent metadata techniques and providing causal consistency, SATURN ensures increased operational capacity and reduced latency of updated data reading operations \cite{bravo2017saturn}.

\subsection{Energy consumption}

In recent years, due to the growth in the power hunger of the servers the Internet based service providing companies like Google, Microsoft, Amazon, etc. are faced with extremely high energy costs. However, the trade-off between energy consumption and consistency is the new challenge with which has gained lots interest among researchers. In order to reduce the energy consumption the consistency must be sacrificed. For instance, Caelus is a device which utilizes both of the causal and eventual consistencies thanks to which its battery life has been extended\cite{Kim2015a}. Yet, with proposition of the Hot \& Cold replication technique and the causal consistency the power consumption has been minimized in the distributed storage system \cite{chihoub2015exploring}.
	
\section{Conclusion}\label{sec.conclusion}
With the emergence of the distributed computations systems as well as the replication process in these systems beside the scalability and availability concepts, the consistency models have been proposed. Our purpose is to concentrate on the consistency models which are proposed in the distributed systems. These models are categorized into to groups data-centric and client-centric by the researchers. In this paper, first the we had and introduction on the main categories of consistency models which are divided into data-centric, client-centric and the hybrid models. Hybrid models are a combination of the different consistency models which are proposed in this paper. Besides, we have shown the contributions of different consistency levels in distributed systems. Subsequently, the traditional, extended, and novel are those consistencies which are proposed based the previously mentioned subcategories consistency models.

Next, the functionality of the proposed consistency models categorized in these subgroups are explained in detail. Consequently, in order to have a full insight of the detailed process performed in the conventional data and client-centric models, we have explained them in mathematical terms.

By providing a new categorization of the consistency models and novel distributed storage systems over time, we found that distributed systems also require consistency. We analyzed the challenges in the distributed systems such as reliability, availability, latency, etc. as well as trade-offs between these challenges and the consistency. We came to the conclusion from this study that by offering a variety of consistencies in distributed systems and their urgent need for consistency, it can be introduced as a service in these systems.


\section*{References}\label{sec.Ref}
	
	\bibliographystyle{elsarticle-num}
	\bibliography{Plain}

\end{document}